\documentclass{article}

\PassOptionsToPackage{pdfborderstyle={/S/U/W 0.3}, hyperfootnotes=false}{hyperref}
\usepackage[utf8]{inputenc}
\usepackage[T1]{fontenc}
\usepackage{microtype}
\usepackage{etoolbox}
\usepackage{paralist}
\usepackage{amsmath,amssymb}
\usepackage{declmath}
\usepackage{varwidth}
\usepackage{listings}
\usepackage{bigfoot} 
\usepackage{makeidx}\makeindex
\usepackage[a4paper]{geometry}
\usepackage{fullpage}
\usepackage{environ}
\usepackage{accsupp}
\usepackage{multicol}
\usepackage{orcidlink}
\usepackage{verbatim,newverbs}
\usepackage[dvipsnames,svgnames]{xcolor}
\usepackage[backend=bibtex,maxbibnames=100]{biblatex}
\usepackage{booktabs}
\usepackage{hyperref}
\usepackage{renderisabelle}

\DeclareRobustCommand\eraseRight{%
  \BeginAccSupp{method=hex,unicode,ActualText=2326}%
  \ensuremath{\rlap{\textsf x}\rhd}%
  \EndAccSupp{}%
}

\DeclareUnicodeCharacter{03BB}{\ensuremath\lambda}
\DeclareUnicodeCharacter{03C0}{\ensuremath\pi}
\DeclareUnicodeCharacter{03C1}{\ensuremath\rho}
\DeclareUnicodeCharacter{03C6}{\ensuremath\varphi}
\DeclareUnicodeCharacter{03C8}{\ensuremath\psi}
\DeclareUnicodeCharacter{21D2}{\ensuremath\Rightarrow}
\DeclareUnicodeCharacter{21E7}{\ensuremath\Uparrow}
\DeclareUnicodeCharacter{21E9}{\ensuremath\Downarrow}
\DeclareUnicodeCharacter{2200}{\ensuremath\forall}
\DeclareUnicodeCharacter{2203}{\ensuremath\exists}
\DeclareUnicodeCharacter{2208}{\ensuremath\in}
\DeclareUnicodeCharacter{2209}{\ensuremath\notin}
\DeclareUnicodeCharacter{2211}{\ensuremath\sum}
\DeclareUnicodeCharacter{2218}{\ensuremath\circ}
\DeclareUnicodeCharacter{2219}{\ensuremath\bullet}
\DeclareUnicodeCharacter{221E}{\ensuremath\infty}
\DeclareUnicodeCharacter{2227}{\ensuremath\land}
\DeclareUnicodeCharacter{2260}{\ensuremath\neq}
\DeclareUnicodeCharacter{2261}{\ensuremath\equiv}
\DeclareUnicodeCharacter{2264}{\ensuremath\leq}
\DeclareUnicodeCharacter{2265}{\ensuremath\geq}
\DeclareUnicodeCharacter{2287}{\ensuremath\supseteq}
\DeclareUnicodeCharacter{2293}{\ensuremath\sqcap}
\DeclareUnicodeCharacter{2294}{\ensuremath\sqcup}
\DeclareUnicodeCharacter{22A4}{\ensuremath\top}
\DeclareUnicodeCharacter{22A5}{\ensuremath\bot}
\DeclareUnicodeCharacter{22C0}{\ensuremath\bigwedge}
\DeclareUnicodeCharacter{2326}{\eraseRight}
\DeclareUnicodeCharacter{27F6}{\ensuremath\longrightarrow}
\DeclareUnicodeCharacter{27F7}{\ensuremath\longleftrightarrow}
\DeclareUnicodeCharacter{27F9}{\ensuremath\Longrightarrow}
\DeclareUnicodeCharacter{290F}{\ensuremath{\relbar\joinrel\longrightarrow}}
\usepackage{newunicodechar}
\newunicodechar{‹}{\textsf\guilsinglleft}
\newunicodechar{›}{\textsf\guilsinglright}
\newunicodechar{∨}{\ensuremath\vee}

\input{macros}

\newcommand\FN@totalidautorefname{footnote}

\DeclareNewFootnote[para]{default}

\AtBeginDocument{
  \ifx\PreviewMacro\undefined\else
    \PreviewEnvironment*{figure}
    \PreviewMacro*\footnote
    \PreviewMacro[!]\isaterm
    \PreviewMacro[!]\isatyp
    \PreviewMacro[!]\isaconst
    \PreviewMacro[!]\isaconstidx
    \PreviewMacro[!]\isaclass
    \PreviewMacro[!]\isaclassidx
    \PreviewMacro[!]\isatop
  \fi}

\bibliography{bounded-operators}

\title{Complex Bounded Operators in Isabelle/HOL%
  \thanks{A short version of this paper appeared at ITP 2026 \cite{itp-shortversion}}}
\author{\orcidlinki{Dominique Unruh}{0000-0001-8965-1931}\\
  \small RWTH Aachen, University of Tartu\\
  \small \emailaddress{unruh-bounded-operators@qis.rwth-aachen.de}
  \and
  \orcidlinki{José Manuel Rodríguez Caballero}{0000-0002-3100-1722} \\
  \small Mathématiques et statistique -- Faculté des sciences et de génie -- Université Laval \\
  \small \emailaddress{jose-manuel.rodriguez-caballero.1@ulaval.ca}
}
\date{}

\begin{document}

\maketitle


\makeatletter

\begin{abstract}
  We present a formalization of bounded operators on complex vector spaces in Isabelle/HOL.
  Our formalization contains material on complex vector spaces (normed spaces, Banach spaces, Hilbert spaces)
  that complements and goes beyond the developments of real vectors spaces in the Isabelle/HOL standard library.
  We define the type of bounded operators between complex vector spaces (\textit{cblinfun}) and develop the theory of
  unitaries, projectors, extension of bounded linear functions (BLT theorem), adjoints, Loewner order, closed subspaces and more.
  For the finite-dimensional case, we provide code generation support by identifying finite-dimensional operators with matrices
  as formalized in the \isasession{Jordan_Normal_Form} AFP entry.
\end{abstract}

\vspace{5mm}

{
  \setlength{\columnseprule}{0.1pt}
  \setlength{\columnsep}{25pt}
  \begin{multicols}{2}
    \@starttoc{toc}
  \end{multicols}
}

\vspace{5mm}

\thispagestyle{empty}

\section{Introduction}

Functional analysis, especially the theory of Hilbert spaces and of operators on these, form an important area in mathematics. In the Isabelle/HOL \cite{isabelle} library \isasession{Complex_Bounded_Operators}~\cite{Complex_Bounded_Operators-AFP}, we formalized a large amount of theorems about complex Hilbert spaces and (bounded) operators on these.

Undoubtedly, these form an important area of mathematics and deserve formalization in a theorem prover for their own sake. However, our foremost motivation for this formalization stems from a specific use case:
We worked with and formalized results related to the theory of quantum programming languages and quantum Hoare logics, and encountered an unfortunate situation: Many results in the literature explicitly restrict themselves to quantum languages with only finite variables.
(By finite variable we mean a program variable with finite domain such as a quantum boolean or qubit, which excludes, e.g., quantum integers.)
While this may be efficient for low-level quantum languages such as assemblers, it puts a restriction on higher-level languages. Those papers which do attempt to formalize infinite variables need infinite-dimensional Hilbert spaces; the resulting mathematics is subtly different from the linear algebra encountered in finite-dimensional quantum mechanics (which works with matrices, essentially).
As a result,  intuition and theorems that apply to finite-dimensional systems only are often used without checking if they hold in the infinite-dimensional case and definitions omit important considerations such as the types of operators (e.g., bounded, trace-class, \dots) and the topologies used in infinite sums and limits.
Sometimes it is not even really clear whether the paper implicitly assumes the finite-dimensional case after all, adding even more uncertainty about what is correct in the infinite-dimensional case.
While we do not claim that the papers\footnote{For example, we encountered the following issues that make it unclear whether the authors considered the mathematical intricacies of infinite-dimensional Hilbert spaces:

\cite{ying12floyd} defines unitaries as operators with $U^\dagger U=1$ which is only correct in the finite-dimensional case. When defining quantum measurements, an infinite sum over operators is used without specifying the topology. When defining the tensor product of countably many finite-dimensional Hilbert spaces, they claim the result has a countable basis. In the definition of the language semantics, more infinite sums occur without specifying the topology. (In the infinite-dimensional case, there are at least five relevant topologies on operators that occur commonly in the context of these papers: operator norm topology, strong operator topology, weak operator topology, trace-norm topology, weak* topology. Therefore not distinguishing them can lead to incorrect results when considering limits and infinite sums.)

\cite{olmedo2019runtime} defines
all mathematical notions (such as linear operators) like one would in the finite-dimensional case (e.g., linear operators instead of bounded operators; no topologies of operators are ever considered),
yet integer-valued quantum variables are used which means the Hilbert spaces must be infinite-dimensional.
They import results about qGCL from \cite{sanders00quantum}
(which has no infinite variables) but use qGCL in a setting with integer variables.

\cite{ying2017invariants} defines
all mathematical notions like one would in the finite-dimensional case (e.g., trace is assumed to be defined on all operators),
yet integer-valued quantum variables are used.
In the semantics of loops, the topology of summation is not specified.

\cite{zhou19applied}
models infinite-dimensional Hilbert spaces, yet claims any operator has a trace (not just trace-class operators).
Figure 2 (the proof system) does not specify a topology for summation.
In the proof of Theorem 3.1, topologies of summation are not mentioned, yet when moving sums in and out of sums and matrix multiplications, different operator topologies would be needed in different proof steps (and no justification is given why the operations preserve limits).

\cite{feng2013model}
\emph{probably} uses infinite-dimensional Hilbert spaces. (Their definition of Hilbert spaces is the infinite-dimensional one, but it is never explicitly mentioned whether the results might implicitly assume finite-dimensional ones).
Yet the preliminaries use linear operators where probably bounded operators would be needed, define trace for all operators (not just trace-class),
and present the spectral theorem that holds (in this form) only for compact operators but claim it holds for all linear operators.
They never discuss operator topologies or convergence.

\cite{zhou21bunched}
seems to use infinite-dimensional Hilbert spaces
(because they import the language definition from \cite{ying12floyd} which has integer variables,
and because Definition 2 contains a sum presumably over $n\in\setN$).
Yet they define density operators as sums of rank-1 operators without specifying the summation topology.
They define the partial trace on values of the form $\sigma\otimes\tau$ (for rank-1 $\sigma,\tau$) and claim it is well-defined by linearity.
(Not true: the $\sigma\otimes\tau$ do not linearly span the space of all operators,
instead the closed span w.r.t.~trace-norm covers all trace-class operators,
so only the \emph{bounded}-linear extension on \emph{trace-class} operators is well-defined.)
In the full version, they claim that the definition of trace is independent of the choice of orthonormal basis
(this is only true for trace-class operators), and they define the SWAP operator without specifying the topology of the sum
(and said sum actually does not converge in the norm-topology which is the most canonical one).

For \cite{yu2019temporal}
we were unable to determine whether the paper was formulated with respect to only finite-dimensional spaces or not.
They explicitly mention that Hilbert spaces can be finite-dimensional or separable.
It is unclear whether this is meant as ``we consider finite-dimensional and separable Hilbert spaces'', i.e., also infinite-dimensional ones,
or as a mathematical claim (which would then be untrue since there exist non-separable Hilbert spaces).
If they consider separable Hilbert spaces, then the definition of the trace as the sum of diagonal entries of the operator is not correct (without prior restriction to trace-class operators).
They import Lemma 2.3 from \cite{li2013termination} and state it without restriction to finite-dimensional Hilbert spaces,
yet it is not true for infinite-dimensional ones (\cite{li2013termination} has a restriction to finite-dimensional spaces in their preliminaries that was not copied over).
Note that \cite{yu2024temporal} by the same author explicitly limits itself to finite-dimensional spaces
but we do not consider \cite{yu2024temporal} to supersede \cite{yu2019temporal} since the latter contains some results the former does not (and vice versa).

Many works, e.g. \cite{ying09predicate}, cite Kraus \cite[Section 3, Theorem 1]{kraus}
for the result that the two common definitions of quantum channels, as completely positive trace-preserving maps and as Kraus operator sums, are equivalent.
It is unclear whether Kraus' book contains a hidden restriction to only separable Hilbert spaces.
(It says ``the basic quantities [\dots] of the theory may be represented mathematically in terms of operators on a (complex, separable) Hilbert space $H$'' in its introduction.
We do not know whether Theorem 1 applies to non-separable spaces.)

All this may sound like nit-picking, but when these mathematical finer
points are not explicitly and carefully considered, it is hard to verify whether the
results are indeed correct.

}
in which we encountered these problems are necessarily wrong, we believe that it is basically impossible to independently verify the results without a background in functional analysis and a thorough re-checking of the definitions and proofs.
This situation is, basically, the archetypal motivation for formal verification and led us to formalizing the present results. An additional motivation is that existing Isabelle/HOL formalizations explicitly restrict results from the literature due to missing formalized mathematical background.
For example, Li et~al.~\cite{liu19formal} formalize the logic from \cite{ying12floyd} in Isabelle/HOL but explicitly restrict themselves to only finite variables compared with the original \cite{ying12floyd}.
Finally, our own efforts at formalizing quantum logics (qrhl-tool \cite{qrhl-tool} and quantum references \cite{Registers-AFP}, see the related work, \autoref{sec:related-work}) both needed infinite-dimensional spaces (and used earlier versions of our library), emphasizing the need for such a formalization.

Notwithstanding this motivation, we stress that our results are not limited to the use case of formalizing quantum mechanics but cover an area of mathematics of high importance in its own right.

When attempting to formalize an area in mathematics (as opposed to one specific theorem) one question is what selection of definitions and theorems to include.
Including the content of a complete graduate-level textbook (such as \cite{conway2013course}) is tempting as it provides a nice guideline of what a true expert in the field considers relevant,
but this was beyond the scope of our work.
Instead, we used the development of qrhl-tool \cite{qrhl-tool} and of the quantum reference formalization \cite{Registers-AFP} as the guiding principle: We included all concepts and theorems about complex Hilbert spaces and bounded operators that turned out to be required there and formalized them.
We believe that this led to a well-rounded selection of supported material, though of course it is always possible to add more.

The next question is: Why did we use the Isabelle/HOL theorem prover, specifically, as opposed to, e.g., Rocq \cite{rocq} or Lean \cite{lean}?
There are, in our opinion, several benefits of Isabelle/HOL for this project:
First, it has very good automation (in particular due to the SMT-solver-based sledgehammer tool \cite{sledgehammer} which automatically shows many otherwise trivial-seeming but very tedious proof steps). This can be a huge time saver.
Second, it already comes with a very good support for mathematics involving topology, continuity, analysis, etc.
While we are not aware of a systematic comparison of the coverage of Isabelle/HOL with, e.g., Mathlib \cite{mathlib} in Lean
or math-comp \cite{mathcomp} in Rocq, it seemed to us that Isabelle's coverage and maturity is best for the project at hand.
Third, Isabelle/HOL is based on higher-order logic (HOL) \cite{logics}, a logic without dependent types or other advanced type-theoretic concepts.\footnote{Though the Isabelle/HOL provides many \emph{derived} constructions, such as induction, as something that is proved outside its trusted core.}
At the first glance, this may sound like a disadvantage, but we consider it beneficial for two reasons:
From a philosophical point of view, we can put higher trust in HOL being sound than in logics with advanced type-theoretic features whose interaction can be subtle and difficult to understand.
And: being the common denominator of many logics, there may be a chance that proof terms from Isabelle/HOL proofs could be automatically translated into the other theorem provers, maybe allowing to reuse our results in other theorem provers, while the other direction is unlikely to be possible.
In the end, of course, the choice is subjective and somewhat boils down to a matter of informed taste.

\subsection{Overview of our contribution}
\label{sec:overview}
The main goal of our library is to develop the most important results about complex Hilbert spaces and complex bounded operators.
The Isabelle/HOL distribution (session \isasession{HOL-Analysis}) already contains rich material about \emph{real} vector spaces, as well as \emph{real} bounded operators (called bounded linear functions, \isatypidx{blinfun}, there), although not to the extent covered in our library.
(See~e.g.~\autoref{fig:counts}.)
In order to make our library more aligned with the existing developments, we tried to follow the overall structure of the existing developments.
In Isabelle/HOL, results on real vector spaces and bounded operators are mostly found in the theories \isathyidx[HOL]{Real_Vector_Spaces}, \isathyidx[HOL-Analysis]{Inner_Product}, and \isathyidx[HOL-Analysis]{Bounded_Linear_Function}.
We use an analogous structure, leading to the following theories:
\begin{compactitem}
\item \isathyidx{Complex_Vector_Spaces}:
  Results about complex vector spaces but not involving inner products. This includes the definition of complex vector spaces, normed complex vector spaces, complex algebras, Banach spaces and more (as type classes);
  definitions of linear, bilinear, bounded linear functions (as predicates);
  \emph{antilinear and sesquilinear functions; finite-dimensional spaces; closed subspaces of a complex vector space and sums thereof;
    conjugate spaces; products of complex vector spaces.}
\item \isathyidx{Complex_Inner_Product}:
  Results about inner product spaces. This includes the definition of inner product spaces (as a type class);
  inner product spaces are normed spaces; \emph{complex Hilbert spaces; orthonormal bases; orthogonal complements; projections; Riesz representation; adjoints}.
\item \isathyidx{Complex_Bounded_Linear_Function}:
  Introduces the type $\alpha\ \cblinfunA\ \beta$ (a.k.a.~\isatyp{cblinfun}) of bounded operators (a.k.a.~bounded linear functions) between two complex normed vector spaces.
  Having a dedicated type (as opposed to a predicate on functions) for bounded operators makes many results easier to state.
  E.g., bounded operators themselves form a normed vector space; hence all results stated for the typeclass \isaclass{complex_normed_vector} automatically apply to bounded operators.
  Results include: The type \isatyp{cblinfun}; instantiation as a normed vector space / Banach space; strong operator topology;
  \emph{adjoints\footnote{%
    Adjoints, projectors, and the Riesz representation theorem were already introduced in \texttt{Complex\_Inner\_Product}. Here, we lift those definitions to the type \texttt{cblinfun} and prove many more properties.}; isomorphisms;
  unitaries, isometries, and partial isometries; bounded operators on products of spaces; images of subspaces under bounded operators; projectors; Loewner order on bounded (square) operators; vectors as degenerate bounded operators; rank 1 operators; Banach--Steinhaus theorem;
  Riesz representation theorem; extension of bounded operators (BLT-theorem).}
\end{compactitem}
This goes beyond the results that the Isabelle/HOL distribution has for real bounded operators.
In the above bullet points, italic font marks material that has no analogue in the existing formalization for real bounded operators.
The material before the $\blacktriangle$, however, is already proven in the real case.
While we cannot directly reuse that material, in most cases the definitions and proofs carry over with only minimal changes.
To make it easy to compare the complex with the real material, to get more unified naming conventions, and to make porting of future additional material on real vector spaces to our library easier, we have split off all material that is exactly analogous to the real vector space theories into theories ending with~\texttt{0}, i.e., \isathyidx{Complex_Vector_Spaces0}, \isathyidx{Complex_Inner_Product0}, \isathyidx{Complex_Bounded_Linear_Function0}.
These theories stand in one-to-one correspondence with the corresponding real ones with all definitions/lemmas in the same order. Definition/lemma names are typically marked just with an extra \texttt{c} or \texttt{complex} in the name, and any deviations (e.g., because a theorem does not hold in the complex case) are explicitly mentioned in comments. Anything beyond this is in the theories without \texttt{0}.

Similarly, we have \isathyidx{Complex_Euclidean_Space0} in one-to-one correspondence to the real \isathy[HOL-Analysis]{Euclidean_Space}, except that we have no theory \texttt{Complex\_Euclidean\_Space} because we needed no results on Euclidean spaces beyond the ones already present in the real case.

Beyond the development above, we have the following additional theories:
\begin{compactitem}
\item 
  \isathyidx{One_Dimensional_Spaces}: We introduce the type-class \isaclassidx{one_dim} of one-dimensional complex vector spaces. In pen-and-paper mathematics, we often implicitly identify any one-dimensional space with $\setC$, making many things notationally much simpler. E.g., the adjoint of \isaterm{ψ} applied to \isaterm{φ}, i.e., \isaterm{ψ* φ} (strictly speaking an operator $\setC\to\setC$) is identified with the inner product \isaterm{ψ ∙⇩C φ} (a complex number). Using the type-class \isaclass{one_dim} and suitable isomorphisms, we can make these implicit identifications explicit, with reasonable notational overhead.
\item
  \isathyidx{Complex_L2}: We introduce the Hilbert space \isatyp{'a ell2}\idxtyp{ell2} of square-summable functions $\isatyp{'a}\to\setC$.
  (For finite \isatyp{'a}, this is the same as the space of $n$-dimensional vectors.)
  This is a very canonical Hilbert space because any Hilbert space with basis \isatyp{'a} is isomorphic to \isatyp{'a ell2}.
  Therefore this specific Hilbert space deserves particular attention in its own theory.
\item
  \isathyidx{Cblinfun_Matrix}:
  For finite \isatyp{'a},\isatyp{'b} (of cardinalities $n,m$), the bounded operators \isatyp{'a ell2 ⇒⇩C⇩L 'b ell2} are essentially $(m\times n)$-matrices.
  The \isasessionidx{Jordan_Normal_Form} Archive of Formal Proofs entry (\cite{Jordan_Normal_Form-AFP}, short JNF\index{JNF|see{\isasession{Jordan_Normal_Form}}}) already contains rich material about matrices.
  Our theory \isathy{Complex_L2} establishes the relationship between finite-dimensional bounded operators and JNF-matrices, and relates operations on operators to corresponding operations on JNF-matrices.
\item
  \isathyidx{Cblinfun_Code}:
  We set up Isabelle's code-generation\index{code generation} support for explicit computations involving finite-dimensional vectors, operators, and subspaces.
  This allows to prove many facts involving concrete finite-dimensional operators by computation (under the additional assumption that Isabelle's code generation is sound).
  JNF already contains code-generation setup for matrices, so in many cases, we can simply reduce our code-generation to JNF using the results from \isathy{Cblinfun_Matrix}.  
\end{compactitem}
Finally, there are a lot of miscellaneous results that were required to prove the results in this library. We collected them in theories in the subdirectory \texttt{extra}; two of them have been included in the Isabelle distribution (\isathyidx[HOL-Library]{Complemented_Lattices} and \isathyidx[HOL-Analysis]{Infinite_Sum}).

\begin{table}[t]
  \centering
\begin{tabular}{llll}
  \textbf{Theory} & \textbf{Lines} & \textbf{Defs} & \textbf{Lemmas} \\
  \midrule
  \isathy{Complex_Vector_Spaces0} & 1271 & 3 & 257 \\
  \isathy{Complex_Vector_Spaces} & 2890 & 32 & 215 \\
  \isathy{Complex_Euclidean_Space0} & 300 & 3 & 37 \\
  \isathy{Complex_Inner_Product0} & 442 & 2 & 71 \\
  \isathy{Complex_Inner_Product} & 2340 & 13 & 134 \\
  \isathy{One_Dimensional_Spaces} & 276 & 1 & 30 \\
  \isathy{Complex_Bounded_Linear_Function0} & 788 & 23 & 59 \\
  \isathy{Complex_Bounded_Linear_Function} & 4524 & 43 & 420 \\
  \isathy{Complex_L2} & 1628 & 27 & 97 \\
  \isathy{Cblinfun_Matrix} & 1467 & 7 & 69 \\
  \isathy{Cblinfun_Code} & 550 & 12 & 33 \\
  \isathy{Cblinfun_Code_Examples} & 66 & 0 & 0 \\
  \texttt{extra/*} & 2216 & 20 & 223 \\
  \midrule
  Total & 18758 &	186 &	1645 \\
\end{tabular}
\caption{Number of (non-blank) lines of code, definitions, and proven facts in our development.
  \texttt{extra/*} summarizes all additional theories proving additional theorems (required by but not directly related to our formalization).
  Theories ending in \texttt{0} are analogous to existing Isabelle/HOL content for real vector spaces,
  the rest is fully new.}
\label{fig:counts}
\end{table}

We would also like to stress one limitation of our work:
Almost all of our results are for \emph{complex} vector spaces only.
This is somewhat unfortunate, because a large number of theorems (but not all) hold with essentially the same proofs for both complex and real numbers.
Isabelle/HOL does not provide a mechanism for stating a theorem in a generic way for both complex and real numbers,
so we can only cover both real and complex numbers by duplicating theorems and proofs.%
\footnote{It is possible to define a type class that captures important axioms that hold for both complex and real numbers, and then prove theorems generically
  in this type class, with the theorems then automatically applying to reals and complex numbers.
  In fact, we did this in theory \isathyidx{Extra_Ordered_Fields} by introducing a class \isaclassidx{nice_ordered_field}
  and generalizing a number of lemmas from the Isabelle/HOL distribution from only reals to both reals and complex numbers.
  This approach, however, quickly reaches its limits:
  For example, if a theorem holds both for complex and real numbers, but with an even slightly different proof, then we cannot prove it for \isaclass{nice_ordered_field}.
  The type class \texttt{RCLike} in Lean's mathlib follows the same approach, with, as far as we can tell, the same drawback.
  And even if we overcome this somehow (e.g., by finding a type class definition that allows a theorem of the form: ``if $P$ holds for reals and complex numbers, then $P$ holds for the type class''),
  we still face the problem that we cannot define type classes such as ``vector space'' based on this real/complex field numbers because their definition needs to hard-code the associated scalar field.
  Locales \cite{locales} may allow to circumvent that problem, but they are considerably harder to use than type classes. We have not explored this approach but it might work.}
In fact, we already had to duplicate a number of definitions and lemmas from the Isabelle/HOL distribution which covers only \emph{real} normed vector spaces etc.

\subsection{How to use this paper?}
This paper attempts to fulfill several purposes at once:
 to inform the community about the new Isabelle formalization,
 to provide an overview of our contributions,
and  to serve as a ``manual'' for people who wish to use the formalization as a library,
e.g., when proving theorems in Isabelle/HOL that rely on our theorems about bounded operators.
Due to the large number of theorems, we cannot rely on a user to read through the source code to see what is available.
Instead, we explain most definitions (of important mathematical concepts) here and very tersely mention what results we have shown about them.
So a user who is looking for a certain result (say, about orthogonal projections) can jump to the section describing them (we provide an index for this purpose),
read up on how they are defined, and see there what facts we describe.
(Every mathematical fact we describe there is annotated with a footnote containing the Isabelle lemma name.)
In addition (especially the check for variants of these facts, or smaller results that may not be mentioned in the text),
they can then use the names of the relevant definitions to identify lemmas applying to their needs using the builtin search features in Isabelle (\isatop{find_theorems[* x*]} command or the UI equivalent \cite[Section 3.4.1]{isabelle-jedit}).
We try to keep the introduction of the mathematical concepts self-contained
but we cannot give more than the tersest explanations since we cover many.
We recommend a standard textbook such as \cite{conway2013course}.

The reader not wanting to actually use our formalization (yet) can ignore these tips and simply read the overview (\autoref{sec:overview})
(and optionally first the Isabelle recap in \autoref{sec:prelims}), and then read some or all of Sections~\ref{sec:cvs}--\ref{sec:code} depending on their interest.

\subsection{Related work}
\label{sec:related-work}

The most direct related work is the existing formalization of \emph{real} normed vector spaces, bounded operators, Hilbert spaces in the Isabelle/HOL distribution \cite{isabelle} (session \isasession{HOL-Analysis}~\cite{session-hol-analysis}).
A number of our results are a direct translation of the results there to \emph{complex} numbers.
However, our contribution goes far beyond what is available there; only the content of the theories whose name ends with \texttt{0} has an analogue there.
However, we also strongly draw on the existing material on (non-topological) vector spaces and on topology from the Isabelle/HOL distribution.

Also in the Isabelle/HOL ecosystem, there is the Archive of Formal Proofs entry \isasessionidx{Jordan_Normal_Form} \cite{Jordan_Normal_Form-AFP} which covers a large number of results in linear algebra.
However, all of these are strictly finite-dimensional. Everything is formulated in terms of matrices indexed by integers (as opposed to arbitrary types).
We do, however, draw on their implementation for our code generation for finite-dimensional vector spaces and operators.

Outside the Isabelle ecosystem, most noteworthy are in the theorem prover Rocq \cite{rocq} the math-comp \cite{mathcomp} library, and in Lean \cite{lean} the mathlib \cite{mathlib} library,
both of which aim to cover general mathematics.

  Mathlib in particular contains a very large amount of formalized mathematics
  and is still being expanded by an active community.
  We attempted to compare to what degree the contributions in our work are also present in mathlib.
  An item-by-item comparison is very difficult because of the large number of theorems in the present work, and in mathlib.
  However, we believe that most results in the present work are actually also available in mathlib,
  though the correspondence is often not obvious because the mathlib results are stated at a very high level of abstraction.
  This means that it can be quite difficult to identify the definition/theorem in mathlib for a given definition/theorem in our work. (Even after perusing mathlib's various overview documents.)

  We cross-checked various theorems/definitions to get a feeling for the coverage
  and were only able to find the corresponding mathlib theorems using AI (Perplexity and Claude); in many cases our result or definition can be derived as a special case of some mathlib concept.
  (An additional challenge is that the AI tends to additionally respond what theorems mathlib should have as opposed to what it actually has.)
  Most theorems exist in mathlib; a small example of a theorem that we did \emph{not} find in mathlib is our fact ``an operator is unitary iff it is an isometry with dense image,''\,\isalemmafn{surj_isometry_is_unitary} \autopageref{page:surj_isometry_is_unitary}.
  However, it was possible to generate a 37-line proof of this fact based on mathlib theorems using Claude Code.
  All in all, we suspect that our formalization does not cover much that is not also covered by mathlib. (Except the treatment of one-dimensional spaces, and, of course, the Isabelle-specific challenges such as the code generation setup in \autoref{sec:code} and
  the trick of overcoming typesystem limitations in the definition of positive operators, \autopageref{page:def-pos-ops}.)
  

Follow-up work:
Several Isabelle/HOL formalizations draw heavily on the results from our development due to the need to work with infinite-dimensional Hilbert spaces and operators thereon:
``The Tensor Product on Hilbert Spaces'' \cite{Hilbert_Space_Tensor_Product-AFP} formalizes tensor products on infinite-dimensional Hilbert spaces and on von Neumann algebras; together with additional material on trace-class/compact operators, different operator topologies, Loewner order.
``Quantum and Classical Registers'' \cite{Registers-AFP} formalizes the notion of quantum references~\cite{references} which are roughly speaking pointers into a complex Hilbert space.
``The Oneway to Hiding Theorem'' \cite{Oneway2Hiding-AFP} formalizes the eponymous theorem \cite{ambainis19semiclassical} from quantum cryptography.
``Kraus Maps'' \cite{Kraus_Maps-AFP} formalizes quantum channels as sums of Kraus operators \cite{kraus}, and their properties. ``Compressed Random Oracles'' \cite{Compressed_Oracles-AFP} formalizes the compressed oracle method \cite{zhandry18recording} from quantum cryptography.

Additionally, our theorem prover qrhl-tool \cite{qrhl-tool} for quantum cryptographic proofs uses Isabelle/HOL as a backend and heavily uses our library.

\section{Preliminaries}
\label{sec:prelims}

\paragraph{Mathematical background.}
We assume that the reader is familiar with basic concepts from linear algebra (vector spaces, matrices, linearity, span, \dots).
We furthermore assume that the reader is familiar with basic concepts from topology (open/closed sets, closures, continuity, convergence, \dots).
Beyond that, we try to introduce all mathematical notions as we go along, albeit very tersely.
However, additional knowledge of functional analysis/operator theory is recommended for additional context.
We recommend, e.g., the textbook \cite{conway2013course}.

\paragraph{Isabelle/HOL background.}
To understand our Isabelle formalization in detail (in particular the proofs), one needs to be familiar with Isabelle/HOL\index{Isabelle/HOL}.
We refer to the tutorial \cite{progprove} and reference manual \cite{isarref} for this.
However, to allow readers unfamiliar with Isabelle to read this paper, we give a compact overview of the relevant concepts in Isabelle.
Isabelle\index{Isabelle} \cite{isabelle} is a theorem prover well-suited for general mathematics; it is instantiated with different logics, HOL\index{HOL}\index{higher order logic|see{HOL}} being the most commonly used one and the one used in our work.\footnote{\emph{Isabelle} refers to the theorem prover itself, while \emph{Isabelle/HOL} refers to the prover running specifically with the HOL logic.
Within the context of this paper, they can be treated as synonyms.} HOL is a classical logic (with excluded middle and axiom of choice) with a simple higher-order type system (no dependent types), for details see \cite{logics}.

In Isabelle, the basic organizational structure is a \emph{theory}\index{theory}: a single file containing definitions, theorems, and configuration.
Theories are grouped into \emph{sessions}\index{session}; for example the contribution of this paper is the session \isasession{Complex_Bounded_Operators}.

A mathematical expression is called a \emph{term}\index{term}.
Isabelle comes with a configurable syntax close to informal mathematics that should be mostly self-explanatory, e.g., \isaterm{∀x::nat. (x+1)⇧2 = x⇧2 + 2*x + 1}.
Terms (and some other mathematical content) are often enclosed in \texttt{‹\dots›} to disambiguate them from the surrounding commands.
Noteworthy are type annotations (here \isaterm{[*x*]::nat} indicating that \isaterm{x} is of type \isatyp{nat}, natural number).
Type annotations are optional; if they are omitted, Isabelle uses type inference to determine the most general type of an expression.
Noteworthy is also a technical quirk: Implications can be written both \isaterm{[*x*]⟶[*y*]} or \isaterm{[*x*]⟹[*y*]}, all-quantifiers can be written both \isaterm{∀x[*.y*]} or \isaterm{⋀x[*.y*]}.
The difference influences how Isabelle uses the theorems but is immaterial for understanding this paper.

At the core of Isabelle are \emph{lemmas}\index{lemma} (a.k.a.~\emph{theorems}\index{theorem}). A lemma introduces a new mathematical fact followed by a (computer-verified) proof. For example \isatop{lemma commute: ‹a + b = b + a›[*sorry*]} (followed by a proof) introduces the commutativity of addition as a lemma; the fact can later be used using the lemma name
\isalemmapre{lemma commute: True by simp}{commute}.
We do not describe the proofs themselves here.

New mathematical definitions can be introduced by the \isatop{definition[*"x=1"*]} command.
For example, \isatop{definition ‹one23 = 123›} introduces a new constant \isatermpre{definition ‹one23 = 123›}{one23} that is by definition equal to \isaterm{123}.
One can also define functions with one or more arguments:
\isatop{definition ‹square x = x*x›}. (When the rhs is a boolean, one can also write \isaterm{[*x*]⟷[*y*]} instead of \isaterm{[*x*]=[*y*]}.)
From then on, one can use the new constant interchangeably with the rhs,
and a fact stating the equality between the two is available as \isalemmapre{definition ‹one23 = 123›}{one23_def} or \isalemmapre{definition ‹square x = x*x›}{square_def}.
Free variables (and type variables) in the statement of a lemma are always implicitly all-quantified.

Note that there is also an alternative definition command \isatop{lift_definition[* const :: ‹'a ell2› is ‹λ_. 1› sorry*]}. For space reasons we do not explain it here, suffices to say that it allows us to give more compact definitions of constants in types that were defined using \isatop{typedef[*x="{True}" sorry*]} (see below) without having to explicitly use the morphisms \texttt{Abs\_\dots} and \texttt{Rep\_\dots} all the time.

Types\index{type} can be either basic types (such as \isatypidx{bool}, natural numbers \isatypidx{nat}, real numbers \isatypidx{real}, etc.), or applications of type constructors as in \isatyp{nat list}\idxtyp{list} (list of natural numbers), \isatyp{real option}\idxtyp{option} (option type -- either \isaterm{None} or \isaterm{Some real_number}), \isatyp{nat ⇒ complex} (functions from naturals to complex numbers), \isatyp{nat × nat} (pairs of natural numbers). Type constructors always come behind their arguments, so it's \isatyp{nat list} and not \isatyppre{typedef list = "{True}" sorry typedef 'a nat = "{True}" sorry}{list nat}!
A type constructor with several arguments has its arguments in parentheses, e.g., \isatyp{(nat, real) prod}\idxtyp{prod} (which is synonymous with \isatyp{nat × real}).
Of course, more complex types can be built by nesting, e.g., \isatyp{nat list option ⇒ (real × real)} for a function optionally taking a list of naturals and returning a pair of reals.

Isabelle supports polymorphism by using type variables\index{type variable}\index{variables!type}.
A type variable is written \isatyp{'a}, \isatyp{'b}, \dots\ (Type variables with more than one letter are also allowed but common style goes towards single-letter type variables.) A type variable means any type can be used in this position. E.g., \isatyp{'a × nat} is a pair of something of not further specified type and a natural number. And \isatyp{'a × 'a} refers to a pair of two values of the same type.
We stress that while \isatyp{'a} is not a specific type, it is fixed through a lemma.
So something of type \isatyp{('a × 'a) list} cannot be, e.g., the list \isatermpre{definition "True=1" for True definition "False=1" for False}{[(1,2), (True, False)]}. (Imagine something like an all-quantification \texttt{∀'a} around the whole lemma, even though this is not actually valid Isabelle-syntax.)
If a type variable should not be a completely arbitrary type, but a type with certain properties, one can constrain it with a ``sort''\index{sort (of a type)} such as \isatyp{'a::ab_group_add} indicating that \isatyp{'a} should range only over types that are Abelian additive groups. A sort can contain several constraints, e.g., \isatyp{'a::{ab_group_add,topological_space}}. Here \isaclass{ab_group_add} and \isaclass{topological_space} are ``type classes''\index{type class} that tell Isabelle what constants are defined for this type and what axioms hold (e.g., \isaclass{ab_group_add} will require that the addition \isaterm{[*x*]+[*y*]} is defined on \isatyp{'a} and that it is commutative).

We can also define new types and type constructors using the \isatop{typedef[*x="{True}" sorry*]} command.
E.g., we can define the type of all non-empty lists:
\isatop{typedef 'a nonempty = ‹{l::'a list. length l > 0}›[*sorry*]}.
(Plus a proof that the type is inhabited.)
Here the rhs specifies a set that describes what values are in the type.
Notice that the type can be parametrized by a type variable, here \isatyp{'a}.
From then on, we can, e.g., write \isatyppre{typedef 'a nonempty = "{True}"sorry}{nat nonempty} for nonempty lists of naturals.
Note that this does not mean that \isatermpre{type_synonym 'a nonempty = "'a list"}{[1,2] :: nat nonempty} is well-typed.
Instead, \isatyppre{typedef 'a nonempty = "{True}"sorry}{nat nonempty} is a \emph{type copy}, i.e., a separate abstract type, with morphisms (bijections) between \isaconstpre{typedef 'a nonempty = "{True}"sorry}{Rep_nonempty} and \isaconstpre{typedef 'a nonempty = "{True}"sorry}{Abs_nonempty} between the abstract type and the representing set \isaterm{{l. length l > 0}}.

Above, we mentioned type classes\index{type class}\index{class!type}.
A type variable can be constrained to belong to a given type class, and concrete types can then be instantiations of that type class.
This means that a general lemma proven for all types of a given type class (e.g., \isatyppre{class semigroup}{'a :: semigroup}) can be applied to any type that instantiates that type (e.g., \isatyp{nat}).
The type class specifies a set of operations that need to be supported on the type (e.g., a group operation), and axioms that must hold (e.g., associativity).
Many classes are predefined but we can define a new one using the \isatop{class[* x*]} command:
\begin{isabelletop}
  class semigroup = fixes op :: ‹'a ⇒ 'a ⇒ 'a›
    assumes assoc: ‹op (op a b) c = op a (op b c)›
\end{isabelletop}
All types instantiating this class (in particular \isatyppre{class semigroup}{'a::semigroup}) will then have some binary operation called \isaconstpre{consts op :: nat}{op}, as well as an associativity theorem \isalemmapre{lemma assoc: True by simp}{assoc}.
Note that classes can also inherit from each other, e.g., an abelian semigroup can be defined as
\begin{isabelletoppre}{class semigroup = fixes op :: ‹'a ⇒ 'a ⇒ 'a›}
  class ab_semigroup = semigroup + assumes comm: ‹op a b = op b a›
\end{isabelletoppre}

One can then instantiate\index{instantiation!type class}\index{type class!instantiation} the type class with a concrete type.
After defining, e.g., \isaconstpre{consts op::nat}{op} as addition on naturals, we can tell Isabelle that naturals form a semigroup:
\begin{isabelletoppre}{class semigroup}
  instance nat :: semigroup[* sorry*]
\end{isabelletoppre}
(Plus a proof.) Also, if a type has type parameters, then the \isatop{instance[*nat::finite sorry*]} command can put constraints on the type classes of the parameters.
For example, \isatoppre{class semigroup}{instance list :: (finite) semigroup[* sorry*]} would mean that \isatyp{'a list} is a semigroup whenever \isatyp{'a} is finite.
Note that there is a variant of the \isatop{instance[*nat::finite sorry*]} command called \isatop{instantiation[*nat::finite begin instance sorry end*]}.
For the purposes of this paper, the difference is immaterial.
More information about type classes and instantiations can be found in~\cite{classes}.

An alternative to type classes are \emph{locales}\index{locale}.
These are an Isabelle-specific concept to group assumptions and constant definitions together.
Describing these is beyond the scope of this intro, see~\cite{locales} for more information.


\section{Complex vector spaces}
\label{sec:cvs}

\paragraph{Type classes.} Before we can even define complex bounded operators,
we need to define complex vector spaces\index{vector space!complex}\index{complex vector space}
and associated notions.
Isabelle/HOL already has a general definition of vector spaces over arbitrary fields, via the locale \isalocaleidx{vector_space}. However, locales lack the ease of use and automation that are available when using type-classes.\footnote{%
  This is the author's \emph{personal} experience; opinions may differ.
  We found that the use of locales leads much more often to hard to comprehend errors and unexpected impossibilities.
  As far as we can tell, this is because locales in many situations have to tweak the Isabelle system
  to make things look right. (Such as making the parser/printer pretend something which is in the kernel a free variable to be a constant.)
  On the other hand, locales are a very powerful tool when used right.
  We chose to try to avoid them in favor of type classes, anyway, as far as feasible.}
We define the class of complex vector spaces by:
\begin{isabelletop}
class complex_vector = scaleC + ab_group_add +
  assumes scaleC_add_right: ‹a *⇩C (x + y) = (a *⇩C x) + (a *⇩C y)›
    and scaleC_add_left: ‹(a + b) *⇩C x = (a *⇩C x) + (b *⇩C x)›
    and scaleC_scaleC[simp]: ‹a *⇩C (b *⇩C x) = (a * b) *⇩C x›
    and scaleC_one[simp]: ‹1 *⇩C x = x›
\end{isabelletop}
\idxclass{complex_vector}\idxclass{scaleC}
This means a type is a complex vector space if it is an abelian additive group (class \isaclass{ab_group_add}\idxclass{ab_group_add}), has an operation \isaconst{scaleC}\idxconst{scaleC} (written as infix \symbolindexmark\scaleC\scaleC) for multiplication with a complex number, and satisfies the specified laws. This is analogous to the existing class \isaclass{real_vector}\idxclass{real_vector}, except that scalar multiplication is complex.

We define many other subclasses of \isaclass{complex_vector}, most importantly \isaclass{complex_algebra}/\isaclass{complex_algebra_1}\idxclass{complex_algebra}\idxclass{complex_algebra_1} adding the axioms of an algebra (with multiplicative unit), \isaclass{complex_normed_vector}\idxclass{complex_normed_vector} requiring the existence of a norm (written \texttt{norm}\idxconst{norm} or $\norm\cdot$) and having a topology induced by this norm, and \isaclass{cbanach}\idxclass{cbanach} for complex Banach spaces (complex normed spaces that are topologically complete; simply defined as the intersection \isatop{class cbanach = complex_normed_vector + complete_space}). A complex vector space is always also a real vector space, so we prove that \isaclass{complex_vector} is a subclass of \isaclass{real_vector} (Isabelle syntax: \isatop{subclass (in complex_vector) real_vector[* sorry*]}), and similarly for all the other type classes.

\paragraph{Subspace, span, linear.} Probably the most important notions related to vector spaces are the notion of a subspace\index{subspace} (closed under addition and scalar multiplication), the span\index{span}, and the notion of a linear function\index{linear function}. 
\begin{quote}
  \isaterm{csubspace :: ('a::complex_vector) set ⇒ bool}\\
  \isaterm{cspan :: ('a::complex_vector) set ⇒ 'a set}\\
  \isaterm{clinear :: ('a::complex_vector ⇒ 'b::complex_vector) ⇒ bool}
\end{quote}
\idxconst{csubspace}\idxconst{cspan}\idxconst{clinear}
We get their definitions for free by instantiating the general theory of vector spaces.
In case of infinite-dimensional normed vector spaces, bounded linear functions%
\index{bounded linear function}\index{linear function!bounded}
(a.k.a.~bounded operators)%
\index{bounded operator}\index{operator!bounded}
are highly important (and the foundation of most material below).
We define the predicate \isaterm{bounded_clinear :: ('a::complex_normed_vector ⇒ 'b::complex_normed_vector) ⇒ bool} by:
\begin{isabelletop}
locale bounded_clinear = clinear f for f +
  assumes bounded: ‹∃K. ∀x. norm (f x) ≤ norm x * K›
\end{isabelletop}
\idxlocale{bounded_clinear}
That is, a bounded linear function is a linear one that additionally satisfies the axiom \texttt{bounded}. (Logically, this locale definition is equivalent to \isatop{definition ‹bounded_clinear f ⟷ clinear f ∧ (∃K. ∀x. norm (f x) ≤ norm x * K)›[* for bounded_clinear*]}.)
We also define bounded bilinear functions%
\index{bounded bilinear function}%
\index{bilinear function!bounded}
similarly (\isaconstidx{bounded_cbilinear}).
(Here ends the material with direct analogue in the theory \isathy[HOL]{Real_Vector_Spaces}.)

\paragraph{Antilinear, sesquilinear, conjugate space.}
In the complex case, we also need antilinear\index{antilinear function} functions. An antilinear function is an additive function with \isaterm{f (c *⇩C x) = cnj c *⇩C f x}.
(The difference to a linear function is the added \isaterm{cnj}, denoting the complex conjugate.)
We define the predicate:
\begin{quote}
  \isaterm{antilinear :: ('a::complex_vector ⇒ 'b::complex_vector) ⇒ bool}
\end{quote}
Similarly, we define \isaconst{bounded_antilinear}, and finally \isaconst{bounded_sesquilinear}\index{sesquilinear} for a bounded function that is antilinear in its first and linear in its second argument.

Most results for linear functions have direct analogues for antilinear ones. However, stating every result twice would lead to a lot of duplication.
Instead, we introduce the notion of a conjugate space.%
\index{conjugate space}
Specifically, for a type \isatyp{'a}, we define a type copy\pagelabel{page:conjugate_space}
\begin{isabelletop}
typedef 'a conjugate_space = ‹UNIV :: 'a set›[* sorry *]
\end{isabelletop}
\idxtyp{conjugate_space}
with morphisms \isaconst{from_conjugate_space}\idxconst{from_conjugate_space} and \isaconst{to_conjugate_space}\idxconst{to_conjugate_space} between \isatyp{'a conjugate_space} and \isatyp{'a}.
We set this type copy up so that most constants (e.g., \isaconst{plus}, \isaterm{0}, \isaconst{norm}) on \isatyp{'a conjugate_space} are defined directly in terms of the corresponding operations on \isatyp{'a}, except \isaconst{scaleC} which is defined by:
\begin{isabelletop}
lift_definition scaleC_conjugate_space
  :: ‹complex ⇒ 'a[* :: complex_vector *] conjugate_space ⇒ 'a conjugate_space›
  is ‹λc x. cnj c *⇩C x›[* sorry *]
\end{isabelletop}
(This means that \isaterm{c *⇩C x} on \isatyp{'a conjugate_space} is defined as \isaterm{cnj c *⇩C x} on \isatyp{'a}, with the morphisms automatically inserted.) With this setup, \isatyp{'a conjugate_space} is a complex (normed) vector space if \isatyp{'a} is, and we show that \isaconst{from_conjugate_space} and \isaconst{to_conjugate_space} are (bounded) antilinear isomorphisms.
Why is this useful?
Given an antilinear function \isaterm{f}, the composition \isaterm{f o from_conjugate_space} (or \isaterm{to_conjugate_space o f}) is linear.
We can now apply existing results for (bounded) linear functions to \isaterm{f o from_conjugate_space} and then derive the corresponding result for \isaterm{f} as a corollary. (An example of this reasoning is \isalemma{antilinear_equal_ket} that gives a criterion for equality of antilinear functions, derived directly from analogous results for linear functions.)

\paragraph{Finite-dimensional spaces.}%
\index{finite-dimensional space}
Most of our results are for all complex vector spaces, irrespective of their dimension.
Some results, however, hold only for finite-dimensional vector spaces (or have fewer side conditions in that case).
To state those compactly, we define the type class \isaclass{cfinite_dim} of finite-dimensional complex vector spaces:
\begin{isabelletop}
[*setup ‹Sign.add_const_constraint (\<^const_name>‹cspan›, SOME \<^typ>‹'a set ⇒ 'a set›)›
*]class cfinite_dim = complex_vector +
  assumes cfinitely_spanned: ‹∃S. finite S ∧ cspan S = UNIV›
\end{isabelletop}
\idxclass{cfinite_dim}
An example of a non-trivial fact that holds only for finite vector spaces is:
\begin{isabelletop}
lemma bounded_clinear_finite_dim[simp]:
  fixes f :: ‹'a::{cfinite_dim,complex_normed_vector} ⇒ 'b[*::complex_normed_vector*]›
  assumes ‹clinear f› shows ‹bounded_clinear f›[* sorry *]
\end{isabelletop}
\idxlemma{bounded_clinear_finite_dim}
That is, any linear function on a finite-dimensional space is bounded.
(This comes in very handy if we want to represent bounded operators as matrices, see \autoref{sec:matrix}.)

Other important facts are \isalemma{finite_cspan_closed} and \isalemma{finite_cspan_complete} stating that the span of a finite set is topologically closed and complete, respectively.
Closed subspaces will play a big role later on, so this result is very useful to omit closedness assumptions in the finite-dimensional case.

In addition to \isaclass{cfinite_dim}, we also define a type class \isaclass{basis_enum}\idxclass{basis_enum}.\pagelabel{page:basis_enum}
\begin{isabelletop}
class basis_enum = complex_vector +
  fixes canonical_basis :: ‹'a list›
  assumes [* "a=a" *][*!\texttt{...}*]
\end{isabelletop}
\idxconst{canonical_basis}
This ensures that every type in this class has an explicit canonical basis \isaconstidx{canonical_basis},%
\index{canonical basis}\index{basis!canonical}
given as a list of vectors.
While every finite dimensional vector space has some finite basis,\isalemmafn{finite_basis} for some purposes (especially code-generation) we may need to be able to explicitly enumerate a canonical basis.
(Analogous to Isabelle/HOL's type-class \isaclassidx{enum}.)
One can think of \isaclass{basis_enum} as a constructive version of \isaclass{cfinite_dim}.

\paragraph{Closed subspaces.}\pagelabel{page:ccsubspace}%
\index{subspace!closed}\index{closed subspace}
Subspaces are a quintessential object when studying vector spaces.
But in the infinite-dimensional case, it is possible that a set is a subspace but not topologically closed.\footnote{%
  For example, in the Hilbert space of square-summable sequences (\isatyp{nat ell2} in the notation of \autoref{sec:ell2}), the set of all sequences with finite support is a subspace but not closed.}
In many situations, we need closed subspaces.
For example, orthogonal projectors (see~\autoref{sec:inner-product}, \autopageref{page:inner.proj}) only exist on closed subspaces.
And for example in Birkhoff--von Neumann quantum logic%
\index{Birkhoff von-Neumann quantum logic}%
\index{quantum logic!Birkhoff von-Neumann}~%
\cite{birkhoff36logic} (and quantum Hoare logics based on that), predicates are closed subspaces.
In light of this, we define the type of closed subspaces.
\begin{isabelletop}
typedef[* (overloaded) *] ('a::‹{complex_vector,topological_space}›) ccsubspace =
    ‹{S::'a set. closed_csubspace S}›[* sorry *]
\end{isabelletop}
\idxtyp{ccsubspace}
That is, if \isatyp{'a} is a complex vector space (with some associated topology, e.g., a normed vector space), then \isatyp{'a ccsubspace} is the type of all closed subspaces of \isatyp{'a}.
This type is abstract; to access \isaterm{S :: 'a ccsubspace} as a set of vectors, we write \isaterm{space_as_set S}.\idxconst{space_as_set}
And given a set \isaterm{S} of vectors, we define \isaterm{ccspan S :: 'a[*::complex_normed_vector*] ccsubspace} to be its closed span.%
\index{closed span}\index{span!closed}

We also define the closed sum of subspaces.%
\index{closed sum!of subspaces}\index{sum!closed, of subspaces}
\begin{isabelletop}
definition closed_sum:: ‹'a[*::{semigroup_add,topological_space}*] set ⇒ 'a set ⇒ 'a set› where
    ‹closed_sum A B = closure (A + B)›
\end{isabelletop}
(Here \isaterm{A + B} is the pointwise sum of two sets.) We also write this as \isaterm{A [*!\symbolindexmarkhighlight{*]+⇩M[*!}*] B}\symbolindexmarkonly\closedSum. If $A$ and $B$ are subspaces, then \isaterm{A +⇩M B} is a closed subspace.\isalemmafn{closed_subspace_closed_sum}
Hence closed subspaces are closed under this operation.
In fact, it turns out that \isaterm{A +⇩M B} is the supremum%
\index{supremum!(of subspaces)}
of $\isaterm A,\isaterm B$ (with respect to set-inclusion.\isalemmafn{closed_sum_is_sup}\fnsep\footnote{%
  Notice that in general, even if $A$ and $B$ are closed subspaces, the pointwise sum $A+B$ is not closed. See, e.g., \cite{mathse-sum-closed}. So it is important to define \isaconst{inf} as the \emph{closed} sum.}
And intersection is the infimum.%
\index{infimum!(of subspaces)}
Thus by defining
\isaconstidx[for \isatyp{ccsubspace}]{sup} and
\isaconstidx[for \isatyp{ccsubspace}]{inf}
on the type \isatyp{ccsubspace} to be \isaconst{closed_sum} and set-intersection, respectively, and defining \isaconstidx[for \isatyp{ccsubspace}]{less_eq} to be the subset-relation, \isatyp{ccsubspace} becomes a lattice\index{lattice}.
The space of all vectors is its maximal element \isaconstidx[for \isatyp{ccsubspace}]{top} a.k.a. \symbolindexmark{\topISA}{\isaterm{⊤}};\isalemmafn{space_as_set_top} the space \isaterm{0}\idxconst[for \isatyp{ccsubspace}]{zero} containing only the vector \isaterm{0} is its minimal element \isaconstidx[for \isatyp{ccsubspace}]{bot}.\isalemmafn{space_as_set_bot}\fnsep\isalemmafn{zero_ccsubspace_def}
In fact, we prove that \isatyp{'a ccsubspace} is even a complete lattice:%
\index{complete lattice!(of subspaces)}%
\index{lattice!complete (of subspaces)}
\begin{isabelletop}
instance ccsubspace :: (complex_normed_vector) complete_lattice[* sorry *]
\end{isabelletop}
(This makes Isabelle automatically recognize \isatyp{'a ccsubspace} as having type class \isaclassidx[for \isatyp{ccsubspace}]{complete_lattice} when \isatyp{'a} is a normed space.)
We will come back to this later when studying Hilbert spaces; there we will even get a complete orthomodular lattice.

For notational convenience, we define \isaconstidx[for \isatyp{ccsubspace}]{plus} on the type \isatyp{ccsubspace} to be the same as \isaconst{sup} since $+$ is a more common notation for the sum of subspaces.

\paragraph{Product spaces.}\pagelabel{page:product-spaces}%
\index{product space!of vector spaces}
In Isabelle/HOL, the product of two types, \isatyp{'a × 'b} a.k.a. \isatyp{('a,'b) prod},
is the type of all pairs from \isatyp{'a} and \isatyp{'b}.
Then the product of two complex (normed) vector spaces is a complex (normed) vector space (partially following existing results for real vector spaces in the Isabelle/HOL theory \isathyidx[HOL-Analysis]{Product_Vector}):
\begin{isabelletop}
instance prod :: (complex_vector, complex_vector) complex_vector[* sorry *]
instance prod :: (complex_normed_vector, complex_normed_vector)
                      complex_normed_vector[* sorry *]
\end{isabelletop}
\idxconst[for \isaclass{complex_vector}]{prod}
\idxconst[for \isaclass{complex_normed_vector}]{prod}
Results going beyond \isathy[HOL-Analysis]{Product_Vector} are the fact that taking the span and taking the Cartesian product commute
\begin{isabelletop}
lemma cspan_Times: ‹cspan (S × T) = cspan S × cspan T›
    if ‹0 ∈ S› and ‹0 ∈ T›[* sorry *]
\end{isabelletop}
\idxlemma{cspan_Times}
and that addition (as a linear function from the product space to the vector space) has operator norm $\sqrt2$.\isalemmafn{onorm_case_prod_plus}


\section{Inner product spaces}
\label{sec:inner-product}

\paragraph{Inner product spaces.}%
\index{inner product space}
An inner product\index{inner product} on a complex vector space \isatyp{'a} is positive non-degenerate sesquilinear function from \isatyp{'a × 'a} to $\setC$ (i.e., antilinear in its first argument). We define the type class of complex inner product spaces, i.e., spaces with an inner product called \isaconstidx{cinner}, also written \isaterm{[*x*]∙⇩C[*y*]}:
\begin{isabelletop}
[* setup ‹Sign.add_const_constraint(\<^const_name>‹dist›, NONE)›
   setup ‹Sign.add_const_constraint (\<^const_name>‹norm›, NONE)›
*]class complex_inner = complex_vector + sgn_div_norm + dist_norm + uniformity_dist + open_uniformity +
  fixes cinner :: ‹'a ⇒ 'a ⇒ complex›
  assumes cinner_commute: ‹cinner x y = cnj (cinner y x)›
    and cinner_add_left: ‹cinner (x + y) z = cinner x z + cinner y z›
    and cinner_scaleC_left [simp]:
            ‹cinner (scaleC r x) y = (cnj r) * (cinner x y)›
    and cinner_ge_zero [simp]: ‹0 ≤ cinner x x›
    and cinner_eq_zero_iff [simp]: ‹cinner x x = 0 ⟷ x = 0›
    and norm_eq_sqrt_cinner: ‹norm x = sqrt (cmod (cinner x x))›
\end{isabelletop}
\idxclass{complex_inner}
(\isaconstidx{cnj} denotes complex conjugation.)
Note that we also require that the norm is defined as $\norm x:=\sqrt{x\cinner x}$.\footnote{%
  The additional standard type classes \isaclass{sgn_div_norm}, \isaclass{dist_norm}, \isaclass{uniformity_dist}, \isaclass{open_uniformity} simply guarantee that certain other common constants used by Isabelle/HOL are defined in a way compatible with the norm.}
This makes an inner product space a normed vector space;
we prove this and declare \isaclass{complex_inner} as a subclass of~\isaclass{complex_normed_vector}.

(Here ends the material with direct analogue in the theory \isathy[HOL-Analysis]{Inner_Product} for real vector spaces.)

Two vectors are orthogonal\index{orthogonal!(vectors)} iff their inner product is $0$ (abbreviated \isaterm{is_orthogonal x y}\idxabbrev{is_orthogonal}).

\paragraph{Hilbert spaces.}%
\index{Hilbert space}
A complex Hilbert space is a complex inner product space that is also topologically complete:
\begin{isabelletop}
  class chilbert_space = complex_inner + complete_space
\end{isabelletop}
\isaclass{chilbert_space}\idxclass{chilbert_space} is also trivially a subclass of \isaclass{cbanach} (Banach spaces), and we prove (i.e., instantiate the type class) that the product of Hilbert spaces is a Hilbert space.

In the following, many lemmas only hold for Hilbert spaces (enforced through type annotations).
For brevity and legibility, we will usually not explicitly state whether a given lemma holds only for Hilbert spaces or for any inner product space.
When in doubt, please refer to the source code of the formalization.

\paragraph{Orthogonal complements.}%
\index{orthogonal complement}%
\index{complement!orthogonal}
The orthogonal complement \isaterm{orthogonal_complement S}\idxconst{orthogonal_complement} of a set~$S$ of vectors consists of all vectors that are orthogonal to all of $S$.
The orthogonal complement is always a closed subspace:
\begin{isabelletop}
lemma orthogonal_complement_closed_subspace[simp]:
    ‹closed_csubspace (orthogonal_complement A)›[* sorry *]
\end{isabelletop}
\idxlemma{orthogonal_complement_closed_subspace}
Moreover, on a Hilbert space, the double orthogonal closure is simply the closed span\isalemmafn{orthogonal_complement_orthogonal_complement_closure_cspan}.
In particular, the orthogonal complement is an involution on the lattice of closed subspaces.
This justifies the definition of \isaterm{- S} (or: \isaterm{uminus S})\idxconst{uminus} on \isatyp{'a ccsubspace} as the orthogonal complement of~\isaterm{S}.
We can then show that \isatyp{'a ccsubspace} is a complete \emph{complemented} lattice\index{complemented lattice}\index{lattice!complemented}, i.e., a lattice with an involution $-x$ such that $x\leq y\iff -y\leq -x$.
Even more, we prove the orthomodular law\index{orthomodular law} $x\leq y\implies x\Sup(-x\Inf y) = y$.
(Here $\Sup$ and $\Inf$ denote binary supremum \isaconst{sup} and infimum \isaconst{inf}, respectively.)
Thus \isatyp{'a ccsubspace} is a complete orthomodular lattice:
\begin{isabelletop}
  instance ccsubspace :: (chilbert_space) complete_orthomodular_lattice[* sorry *]
\end{isabelletop}
Important laws such as De Morgan's laws%
\index{De Morgan's laws!for subspaces}
(\isaterm{- (x ⊔ y) = - x ⊓ - y})\isalemmafn{compl_sup}\fnsep\isalemmafn{compl_inf} follow directly because they hold for all orthomodular lattices.
Note that orthomodular lattices are weaker than Boolean algebras%
\index{Boolean algebra}
(class \isaclass{boolean_algebra}\idxclass{boolean_algebra}), the lattice structure describing classical logic predicates, and instead describe well the lattice of quantum predicates represented by closed subspaces of a Hilbert space (Birkhoff-von-Neumann quantum logic%
\index{Birkhoff von-Neumann quantum logic}%
\index{quantum logic!Birkhoff von-Neumann}~%
\cite{birkhoff36logic}).

\paragraph{Orthogonal projections.}\pagelabel{page:inner.proj}%
\index{orthogonal projection}%
\index{projection|see{orthogonal projection}}%
\index{projector|see{orthogonal projection}}
We turn our attention to orthogonal projections\index{projection!orthogonal}\index{orthogonal projection}; projections or projectors for short.
The easiest (and probably most well-known) way to define a projection is as a self-adjoint idempotent bounded operator ($\adj P=P$, $PP=P$).
This is somewhat limited in its generality because it only allows us to define projections onto closed subspaces (and not, e.g., onto a ball).
Instead, we follow the more general approach presented in \cite{conway2013course} (and recover the well-known definition as a special case in \autoref{sec:complex-b-o}, \autopageref{page:is_Proj_algebraic}).
That is, we say $\pi$ is a projection onto a set $M$ iff $\pi(x)$ is a point in $M$ closest to $x$.
\begin{isabelletop}
definition is_projection_on::‹('a ⇒ 'a) ⇒ [*(*]'a[*::metric_space)*] set ⇒ bool› where
  ‹is_projection_on π M ⟷
             (∀h. is_arg_min (λx. dist x h) (λx. x ∈ M) (π h))›
\end{isabelletop}
This definition is general enough to make sense on any metric space (\isaconstidx{dist} refers to the metric).
In general, the projection onto $M$ may not be unique, but it is on an inner product space when $M$ is convex.\isalemmafn{is_projection_on_unique}
Furthermore, the projection may not always exist, but it does on a Hilbert space when $M$ is convex, closed, and non-empty (e.g., if it is a closed subspace).\isalemmafn{is_projection_on_exists}
For convenience, we define \isaterm{projection M} to be an arbitrary projection onto $M$ (if one exists):
\begin{isabelletop}
definition ‹projection M = (SOME π. is_projection_on π M)›[* for projection *]
\end{isabelletop}
\idxconst{projection}
We show some elementary facts about projections: They are idempotent,\isalemmafn{is_projection_on_idem} their image is $M$,\isalemmafn{is_projection_on_in_image} they are norm-reducing functions\isalemmafn{is_projection_on_reduces_norm} and bounded linear.\isalemmafn{is_projection_on_bounded_clinear}

More interestingly: The kernel of a projection is \isaterm{orthogonal_complement M}.\isalemmafn{is_projection_on_kernel_is_orthogonal_complement}
If $P$ is a projection onto $M$, then $1-P$ is a projection onto \isaterm{orthogonal_complement M}.\isalemmafn{projection_on_orthogonal_complement}
The projection onto a one-dimensional space \isaterm{cspan {t}} maps $x$ to $\frac{\isaterm{t ∙⇩C x}}{\isaterm{t ∙⇩C t}}\isaterm{[*x*] *⇩C t}$.\isalemmafn{is_projection_on_rank1}
If $M,N$ are orthogonal with projectors $P_M,P_N$, the projector onto $M+N$ is $P_M+P_N$.\isalemmafn{projection_plus}
(In all lemmas we assume that $M,N$ are closed subspaces.)

More results on projectors will be presented in \autoref{sec:complex-b-o}, \autopageref{page:cbo.proj}.

\paragraph{Orthonormal bases.}\pagelabel{page:inner:onb}
We define orthogonal sets of vectors%
\index{orthogonal set of vectors}
and orthonormal bases (ONBs):%
\index{orthonormal basis}%
\index{basis!orthonormal}%
\index{ONB|see{orthonormal basis}}
\begin{isabelletop}
definition is_ortho_set :: ‹'a::complex_inner set ⇒ bool› where
  ‹is_ortho_set S ⟷ (∀x∈S. ∀y∈S. x ≠ y ⟶ (x ∙⇩C y) = 0) ∧ 0 ∉ S›
definition is_onb :: ‹'a::complex_inner set ⇒ bool› where
  ‹is_onb E ⟷ is_ortho_set E ∧ (∀b∈E. norm b = 1) ∧ ccspan E = ⊤›
\end{isabelletop}
\idxconst{is_ortho_set}\idxconst{is_onb}
That is, $E$ is an ONB (\isaterm{is_onb E}) iff it consists of mutually orthogonal vectors of norm 1 and its \emph{closed} span is the whole space (\isaconstidx{top} a.k.a. \symbolindexmark{\topISA}{\isaterm{⊤}}).

We prove that every Hilbert space has an ONB; in fact, we prove the stronger statement that every orthonormal set of vectors can be extended to an ONB:
\begin{isabelletop}
lemma orthonormal_basis_exists: 
  assumes ‹is_ortho_set S› and ‹⋀x. x∈S ⟹ norm x = 1›
  shows ‹∃B. B ⊇ S ∧ is_onb B›[* sorry *]
\end{isabelletop}
\idxlemma{orthonormal_basis_exists}
For convenience, we define \isaterm{some_chilbert_basis :: ('a::chilbert_space) set} to refer to an arbitrary but fixed Hilbert space basis (which always exists by the preceding lemma).
In \autoref{sec:complex-b-o}, \autopageref{page:bo:onb}, we also show that all ONBs have the same cardinality.

Given an ONB \isaterm{B}, we can represent any vector \isaterm{ψ} as the (infinite) linear combination \isaterm{∑⇩∞b∈B. (b ∙⇩C ψ) *⇩C b} of basis vectors.\isalemmafn{basis_projections_reconstruct} And the squared coefficients of this linear combination add up to the squared norm (\isaterm{(∑⇩∞b∈B. (cmod (b ∙⇩C ψ))⇧2) = (norm ψ)⇧2}; Parseval identity).\isalemmafn{parseval_identity}

\paragraph{Riesz representation theorem.}\pagelabel{page:riesz-rep}%
\index{Riesz representation theorem}%
\index{Riesz--Frechet representation theorem|see{Riesz representation theorem}}
A central theorem in functional analysis is the Riesz (or Riesz--Frechet) representation theorem. It states that every bounded linear functional on a Hilbert space can be uniquely\isalemmafn{riesz_representation_unique} represented as an inner product with a fixed vector:
\begin{isabelletop}
lemma riesz_representation_existence:
  fixes f :: ‹'a[*::chilbert_space*] ⇒ complex›
  assumes ‹bounded_clinear f›
  shows ‹∃t. ∀x. f x = t ∙⇩C x›[* sorry *]
\end{isabelletop}
This lemma is central for proving the existence of adjoints below.
We will revisit this theorem in \autoref{sec:complex-b-o}.

\paragraph{Adjoints.}\index{adjoint}
If $F$ is a map between two inner product spaces, it has adjoint $G$ iff $F(x)\cinner y = x\cinner G(x)$ for all $x,y$.
\begin{isabelletop}
definition ‹is_cadjoint F G ⟷ (∀x y. (F x ∙⇩C y) = (x ∙⇩C G y))›[* for is_cadjoint *]
\end{isabelletop}
\idxconst{is_cadjoint}
Adjoints are arguably one of the most important basic constructions in operator theory; and for bounded operators on Hilbert spaces, adjoints always exist:
\begin{isabelletop}
lemma cadjoint_exists:
  assumes ‹bounded_clinear G› shows ‹∃F. is_cadjoint F G›[* sorry *]
\end{isabelletop}
\idxlemma{cadjoint_exists}
And adjoints are unique.\isalemmafn{is_cadjoint_unique}
We will revisit adjoints in \autoref{sec:complex-b-o}, \autopageref{page:bo:adj}.

\paragraph{Finite dimensional spaces (continued).} Recall the type class \isaclass{basis_enum} for finite-dimensional spaces with canonical basis (\autopageref{page:basis_enum}). In inner product spaces, it makes sense to additionally require that the \isaconst{canonical_basis} is an orthonormal basis.
\begin{isabelletop}
[* setup ‹Sign.add_const_constraint (\<^const_name>‹is_ortho_set›, SOME \<^typ>‹'a set ⇒ bool›)›
*]class onb_enum = basis_enum + complex_inner +
  assumes is_orthonormal: ‹is_ortho_set (set canonical_basis)›
      and is_normal: ‹⋀x. x ∈ (set canonical_basis) ⟹ norm x = 1›
\end{isabelletop}
(We do not write the more compact \isaterm{is_onb (set canonical_basis)} here because that would lead to the additional proof obligation \isaterm{ccspan (set canonical_basis) = top} when doing a type class instantiation.
This proof obligation would be redundant since it is already implied by \isaclass{basis_enum}.)

Since every finite-dimensional space is complete,\isalemmafn{finite_cspan_complete} we can also show that \isaclass{onb_enum} is a subclass of \isaclass{chilbert_space}. (Which means that a type will be of class \isaclass{chilbert_space} automatically once we instantiate it as \isaclass{onb_enum}.)

\paragraph{Conjugate spaces (continued).}%
\index{conjugate space!(of inner product / Hilbert space)}
We revisit the conjugate spaces (see \autopageref{page:conjugate_space}) and define the inner product on \isatyp{'a conjugate_space} as the swapped inner product on \isatyp{'a}.
\begin{isabelletop}
lift_definition cinner_conjugate_space
  :: ‹'a[*::complex_inner*] conjugate_space ⇒ 'a conjugate_space ⇒ complex›
  is ‹λx y. cinner y x›[* sorry *]
\end{isabelletop}
Then we can show that \isatyp{'a conjugate_space} has type classes \isaclass{complex_inner} and \isaclass{chilbert_space} if \isatyp{'a} has. (Before, we showed only preservation of \isaclass{complex_vector} and \isaclass{complex_normed_vector}.)

\paragraph{Product spaces (continued).}\pagelabel{page:product-spacces.inner}%
\index{product space!(of inner product / Hilbert space)}
In \autoref{page:product-spaces}, we showed that the product of two complex vector spaces is a complex vector space. We show that this is the case also for inner product and Hilbert spaces (with a suitably defined inner product):
\begin{isabelletop}
instance prod :: (chilbert_space, chilbert_space) chilbert_space[* sorry *]
instance prod :: (complex_inner, complex_inner) complex_inner[* sorry *]
\end{isabelletop}
\idxconst[for \isaclass{complex_inner}]{prod}
\idxconst[for \isaclass{chilbert_space}]{prod}


\section{One-dimensional spaces}%
\label{sec:one-dim}%
\index{one-dimensional space}

One-dimensional complex vector spaces take a special role in many calculations.
A one-dimensional vector space is isomorphic to $\setC$, and as such, often implicitly identified with~$\setC$.

A good example is the formula $\psi\adj\psi$ for a vector $\psi\in\calH$, which is, e.g., used to describe the projector onto $\psi$.
This formula is interpreted as follows:
$\psi$ is canonically identified with the bounded operator $\setC\to\calH$, $c\mapsto c\psi$.
(Note here the first implicit occurrence of the one-dimensional space $\setC$.)
Then $\adj\psi$, its adjoint, is a bounded operator $\calH\to\setC$.
Composing it with $\psi$ gives us a bounded operator $\psi\adj\psi:\calH\to\calH$.

Similarly, $\adj\psi\phi$ would be interpreted as:
$\psi$ and $\phi$ are identified with bounded operators $\setC\to\calH$.
Then $\adj\psi\phi$ is the product of two bounded operators, giving a bounded operator $\setC\to\setC$.
$\setC\to\setC$ is a space of bounded operators between one-dimensional spaces, hence itself one-dimensional.
So we can identify it with $\setC$, i.e., $\adj\psi\phi\in\setC$, and in fact $\adj\psi\phi=\psi\ \cinner\ \phi$.%
\footnote{Alternatively, we can read it as an application of $\adj\psi$ to $\phi$.
  This will also give us $\adj\psi\phi=\psi\ \cinner\ \phi$ in the end.
  The point of this section is that all those readings should, in the end, provably give the same results.}
Because of this, in math we often just write $\psi^* \phi$ for the inner product.

To make such interpretations formal, we need two things:
\begin{compactitem}
\item An explicit definition of the identification between $\calH$ and $\setC\to\calH$ (possibly with $\setC$ replaced by other one-dimensional spaces in a canonical way).
\item An explicit definition of the identification between any two one-dimensional spaces. (And laws for reasoning about this identification.)
\end{compactitem}
The first of these is easy and will be done in \autoref{sec:complex-b-o}, \autopageref{page:vectors-as-ops}, constant \isaconst{vector_to_cblinfun}.
The second is the goal of this section, the definition of an isomorphism \isaconst{one_dim_iso} between any two types \isatyp{'a},\isatyp{'b} that are one-dimensional complex vector spaces.

The first step is the definition of the type class \isaclass{one_dim} that captures all one-dimensional vector spaces.
A first idea would be to simply define it as a subclass of \isaclass{complex_vector},
with the only axiom \isaterm{cdim UNIV = 1}, i.e., that the space has dimension~$1$.
However, this has two disadvantages:
\begin{compactitem}
\item It will actually not be possible to canonically define the isomorphism \isaconst{one_dim_iso} between two one-dimensional spaces: There are many possible isomorphisms between one-dimensional spaces, and we need to somehow identify a canonical one.
\item If the space is additionally, e.g., a normed vector space, there will be no guarantee that the norm is compatible with this isomorphism. For example, we could have \isaterm{norm 1 ≠ 1} where the first \isaterm{1} is the vector identified with $1\in\setC$. And similarly for the inner product.
\end{compactitem}

To solve this, it is necessary to have a canonically identified element \isaterm{1} that gets identified with the \isaterm{1} from other one-dimensional spaces.
In other words, we require that the canonical basis contains just one element \isaterm{1}.
This leads to the following definition:
\begin{isabelletop}
class one_dim = onb_enum + one + times + inverse +
  assumes one_dim_canonical_basis[simp]: ‹canonical_basis = [1]›
  assumes one_dim_prod_scale1: ‹(a *⇩C 1) * (b *⇩C 1) = (a * b) *⇩C 1›
  assumes divide_inverse: ‹x / y = x * inverse y›
  assumes one_dim_inverse: ‹inverse (a *⇩C 1) = inverse a *⇩C 1›
\end{isabelletop}
Note that this is a subclass of \isaclass{onb_enum}, which means it is required to have a \isaterm{canonical_basis} that is orthonormal.
And here we require that this basis contains only the constant \isaterm{1} (axiom \isalemma{one_dim_canonical_basis}).
This already fully determines the norm (in particular \isaterm{norm 1 = 1}\,\isalemmafn{norm_one}) and fully determines the inner product \isaterm{[*x*]∙⇩C[*y*]}.
The remaining three axioms make sure that instances of \isaclass{one_dim} have definitions of \isaterm{[*x*]*[*y*]}, \isaterm{[*x*]/[*y*]} and \isaconst{inverse} that are compatible with those in \isatyp{complex} ($\setC$).

The complex numbers are the most obvious example of a one-dimensional space: \isatop{instance complex :: one_dim[*sorry*]}.

Now we can define the canonical isomorphism \isaconst{one_dim_iso} to be one that maps \isaterm{1} to \isaterm{1}.
(Since \isaterm{[1]} is a basis, this uniquely determines it.)
The actual definition is a bit different:
\begin{isabelletop}
definition ‹one_dim_iso a = of_complex (1 ∙⇩C a)›[* for one_dim_iso *]
\end{isabelletop}
We use this definition because it is technically easier to use.
But we show that it is indeed the isomorphism mapping \isaterm{1} to \isaterm{1}.\isalemmafn{one_dim_iso_of_one}

Then \isaconst{one_dim_iso} is indeed a canonical identification between types in the sense that \isaterm{(one_dim_iso :: 'a[*::one_dim*] ⇒ 'a) = id}\isalemmafn{one_dim_iso_id} (a type is identified with itself via the identity) and \isaterm{one_dim_iso (one_dim_iso x) = one_dim_iso x}\isalemmafn{one_dim_iso_idem} (if we identify \isaterm{x :: 'a} with \isaterm{y :: 'b}, and identify \isaterm{y} with \isaterm{z :: 'c}, then we identify \isaterm{x} with \isaterm{z} as well).

In addition, we prove numerous simple simplification rules that use that most operations are in some way compatible with \isaconst{one_dim_iso}, e.g., \isaterm{norm (one_dim_iso x) = norm x}.\isalemmafn{norm_one_dim_iso}

The theory \isathy{One_Dimensional_Spaces} contains the basic definitions and the properties of \isaconst{one_dim_iso} related to normed / inner product spaces.
The additional properties related to bounded operators and the space \isatyp{ell2} (see Sections~\ref{sec:complex-b-o} and~\ref{sec:ell2}) are given in \isathy{Complex_Bounded_Linear_Function} and \isathy{Complex_L2}, respectively, to avoid circular dependencies.
However, we describe them in this section for self-containment.

In \isathy{Complex_Bounded_Linear_Function} (\autoref{sec:complex-b-o}), we introduce the type \isatyp{('a, 'b) cblinfun} of bounded operators (bounded linear functions) mapping \isatyp{'a} to \isatyp{'b}. We show that the type of bounded operators from a one-dimensional space to a one-dimensional space is again one-dimensional:
\begin{isabelletop}
[* class one_dim = one_dim
class complex_inner = complex_inner
*]instantiation cblinfun :: (one_dim, one_dim) complex_inner begin [*!\dots*][*
instance sorry
end*]
\end{isabelletop}
This allows us, for example, to actually state (and prove), that $\adj\psi\phi=\psi\ \cinner\ \phi$
up to canonical isomorphisms as in our example above, namely:
\begin{isabelletop}
lemma
  ‹one_dim_iso ((vector_to_cblinfun ψ)* o⇩C⇩L vector_to_cblinfun φ) = ψ ∙⇩C φ›
  by simp
\end{isabelletop}

We prove a number of simplification rules for one-dimensional bounded operators. For example, composition of operators, application of operators to vectors, multiplication (as defined by the type class \isaclass{one_dim}) and scaling of a vector coincide up to \isaclass{one_dim}.\isalemmafn{one_dim_cblinfun_compose_is_times}\fnsep\isalemmafn{one_dim_cblinfun_apply_is_times}\fnsep\isalemmafn{scaleC_one_dim_is_times} And the identity operator and \isaterm{1} coincide,\isalemmafn{id_cblinfun_eq_1} and the adjoint and complex conjugation,\isalemmafn{one_dim_iso_adjoint_complex} and \isaterm{[*x*]≤[*y*]} as defined for bounded operators matches the one for complex numbers.\isalemmafn{one_dim_loewner_order}

In \isathy{Complex_L2} (\autoref{sec:ell2}), we define the type \isatyp{'a ell2} of square-summable sequences \isatyp{'a ⇒ complex} (for finite \isatyp{'a} of size $n$, this is essentially the space $\setC^{n}$).
As we elaborate there, the type \isatyp{'a ell2} has a canonical basis formed of vectors called \isaterm{ket x} where \isaterm{x} ranges over \isatyp{'a}.
Thus, if \isatyp{'a} is a singleton type (type class \isaclassidx{CARD_1}, its single element is \isaterm{undefined}), we could define \isaterm{canonical_basis = [undefined]} and then prove:
\begin{isabelletop}
[* class one_dim *]instance ell2 :: (CARD_1) one_dim[* sorry *]
\end{isabelletop}
Unfortunately, this is rejected by Isabelle due to conflicts with other instantiations
even though it is provable.%
\footnote{Isabelle puts certain global conditions on the allowed type class instantiations in order to make the type inference unique.
  In this case, we would have both \isatop{[*class basis_enum*]instance ell2 :: (CARD_1) basis_enum[* sorry *]} (as a logical consequence of this instantiation) and \isatop{[*instance *]ell2 :: (enum) basis_enum[* sorry *]} (due to a different instantiation).
  Then when type inference encounters a type \isatyp{'a ell2} which, due to its context would be required to be of class \isaclass{basis_enum}, the type inference would not know whether to assign the type constraint \isaclass{CARD_1} or \isaclass{enum} to \isatyp{'a}.}
We avoid this by instantiating \isatyp{ell2} instead as:
\begin{isabelletop}
instance ell2 :: (‹{enum,CARD_1}›) one_dim[* sorry *]
\end{isabelletop}
Thus, in order for \isatyp{'a ell2} to be recognized as a one-dimensional space, we need to additionally instantiate \isatyp{'a} as \isaclass{enum}.
Fortunately, in most cases this is not a big problem since the class \isaclass{enum} simply requires us to provide an enumeration of all elements in the type \isatyp{'a}; in a singleton type this is very easy.

We can then show that \isaterm{ket x = 1}.\isalemmafn{ket_CARD_1_is_1}
(As there is only one \isaterm{x :: 'a}, this holds for ``all'' \isaterm{x}.)


\section{Complex Bounded Operators}
\label{sec:complex-b-o}

\paragraph{The type of bounded operators.}%
\index{bounded operators}%
\index{operators!bounded}
The central focus of this library are complex bounded operators a.k.a.~bounded linear functions%
\index{bounded linear functions}%
\index{linear functions!bounded}
over complex normed vector spaces.
We already have a predicate \isaconst{bounded_clinear} that recognizes bounded operators but working with an explicit predicate can be cumbersome: When quantifying over bounded operators, one always needs to explicitly add the predicate. The fact that the bounded operators in turn form a complex normed space is harder to state (we cannot simply instantiate the type class \isaclass{complex_normed_vector}). Reasoning about topological properties (such as convergence of sequences of operators) becomes harder in Isabelle/HOL if we cannot define the topology on the whole space. To avoid all these problems, we define the type of complex bounded operators:\footnote{%
  We prefer the term ``bounded operator'' over ``bounded linear function'' in this text. However, we still named the type \isatyp{cblinfun} (complex bounded linear function) and named the theory file \isathy{Complex_Bounded_Linear_Function}.
  We made this design choice to stay close to the naming conventions of the existing results in the Isabelle/HOL distribution about \emph{real} bounded operators.}
\begin{isabelletop}
typedef[* (overloaded)*] ('a, 'b) cblinfun = ‹{f[*::'a::complex_normed_vector⇒'b::complex_normed_vector*]. bounded_clinear f}›[* sorry *]
\end{isabelletop}
\idxtyp{cblinfun}
Instead of \isatyp{('a,'b) cblinfun}, we can also write \symbolindexmarkonly\cblinfunA\isatyp{'a [*!\symbolindexmarkhighlight{*]⇒⇩C⇩L[*!}*] 'b}.
Given a bounded operator \isaterm{f :: 'a ⇒⇩C⇩L 'b}, we can apply%
\index{application!of bounded operators}%
\index{bounded operator!application}
it to a vector \isaterm{x} by writing \isaterm{cblinfun_apply f x}\idxconst{cblinfun_apply} or \isaterm{f [*!\symbolindexmarkhighlight{*]*⇩V[*!}*] x}\symbolindexmarkonly\cbapplyISA.
(\isaterm{f x} is also recognized in many situations due to Isabelle's coercion mechanism.)
Given a function \isaterm{f :: 'a ⇒ 'b} that satisfies \isaterm{bounded_clinear f}, we write \isaterm{CBlinfun f}\idxconst{CBlinfun} to convert it to type \isatyp{'a ⇒⇩C⇩L 'b}.
And the composition%
\index{composition!of bounded operators}%
\index{bounded operator!composition}
(a.k.a.~product)%
\index{product!of bounded operators}%
\index{bounded operator!product}
of two bounded operators is \isaterm{cblinfun_compose A B}\idxconst{cblinfun_compose}, also written \symbolindexmarkonly\oCL{\isaterm{A [*!\symbolindexmarkhighlight{*]o⇩C⇩L[*!}*] B}}.
The identity\index{identity!(bounded operator)}
is \isaterm{id_cblinfun :: 'a ⇒⇩C⇩L 'a[*::complex_normed_vector*]}\idxconst{id_cblinfun}.

The norm\index{norm!of bounded operator} of a bounded operator $A$ is usually defined to be the operator norm,%
\index{operator norm}\index{norm!operator}
i.e., the least~$b$ such that $\norm{A\psi}\leq b\norm\psi$ for all $\psi$.
We follow this convention and define \isaconstidx[for \isatyp{cblinfun}]{norm} on \isatyp{'a ⇒⇩C⇩L 'b} as the operator norm:
\begin{isabelletop}
lift_definition norm_cblinfun :: ‹'a[*::complex_normed_vector*] ⇒⇩C⇩L 'b[*::complex_normed_vector*] ⇒ real› is onorm[* sorry *]
\end{isabelletop}
With this norm, the bounded operators form a complex normed vector space, i.e., we instantiate the class \isaclass{complex_normed_vector}.
When the codomain \isatyp{'b} is even a Banach space, so is the space of
bounded operators.
And if domain and codomain are finite-dimensional, so is the space of bounded operators.
\begin{isabelletop}
instance cblinfun :: (complex_normed_vector, complex_normed_vector)
                         complex_normed_vector[* sorry *]
instance cblinfun :: (complex_normed_vector, cbanach) cbanach[* sorry *]
instance cblinfun :: (‹{cfinite_dim,complex_normed_vector}›,
                      ‹{cfinite_dim,complex_normed_vector}›) cfinite_dim[* sorry *]
\end{isabelletop}
Since \isaclass{complex_normed_vector} in turn defines a topology (the topology induced by the norm),%
\index{topology!induced by norm}
this makes \isatyp{'a ⇒⇩C⇩L 'b} into a topological vector space.
There are many other possible topologies on bounded operators.\footnote{%
  Isabelle's logic forces us to pick one specific norm and topology when instantiating a type class.
    Sometimes there is just one meaningful one or, like here, one that is clearly the most canonical one.
    We can still define other norms and topologies as regular Isabelle constants, but these will be less well integrated.}
For example, we also define the strong operator topology%
\index{strong operator topology}
\index{operator topology!strong}
\index{topology!strong operator}
(\isaconstidx{cstrong_operator_topology}).\footnote{%
  We do not use or explore it further in this development, we mainly included it because it is also defined for real bounded operators in the existing Isabelle/HOL development.}

(Here ends the material with direct analogue in the theory \isathy[HOL-Analysis]{Bounded_Linear_Function} for real bounded operators.)

In the remainder of this theory, we state many results relative to the type \isatyp{'a ⇒⇩C⇩L 'b}.
It would be possible to state them also using the predicate \isaconst{bounded_clinear} but the theorems would become unwieldy.
In the cases where theorems about functions satisfying \isaconst{bounded_clinear} are needed instead, one has to derive them as corollaries of the theorems given here.
(Either by using the morphism \isaconst{cblinfun_apply} and \isaconst{CBlinfun} directly, or by using Isabelle's ``transfer'' mechanism \cite{transfer} for which we have included the relevant setup.)

\paragraph{Miscellaneous.}
The list of all results is too long to state here (and includes many minor but useful facts).
Some important useful facts are that bounded operators $A,B$ are equal if $\psi\cinner A\psi=\psi\cinner B\psi$ for all $\psi$,\isalemmafn{cblinfun_cinner_eqI} that to bound the norm of an operator it is sufficient to bound $\norm{A\psi}$ on a dense set of $\psi$,\isalemmafn{norm_cblinfun_bound_dense} various lemmas about infinite sums of bounded operators, the summability of the Cauchy product of two sequences of vectors,\isalemmafn{Cauchy_cinner_product_infsum} and lemmas relating real and complex bounded operators which sometimes allows us to lift existing theorems about real operators to complex operators (via the morphism \isaconstidx{blinfun_of_cblinfun}).

\paragraph{Images.}%
\index{image!of operator}
Recall that we introduced the type \isatyp{ccsubspace} of closed subspaces (\autopageref{page:ccsubspace}).
It is natural to consider the image \symbolindexmarkonly\cblinfunImage\isaterm{A [*!\symbolindexmarkhighlight{*]*⇩S[*!}*] S} (or \isaterm{cblinfun_image A S}) of a closed subspace \isaterm{S} under a bounded operator \isaterm{A}.
However, in general, that image is not necessarily closed itself\footnote{%
  For example, if $e_i$ ($i\in\setN$) form an orthonormal basis of a Hilbert space $\calH$, and $A$ is the bounded operator that maps $e_i$ to $e_i/i$, then we easily see that the image $I$ of $A$ contains all $e_i$.
  Hence the closure of $I$ is $\calH$.
  But the image $I$ itself is not $\calH$:  $\sum_i e_i/i\in\calH$ is not in $I$.
  So $I$ is not closed.}, but we would like \isaterm{A *⇩S S} to have type \isatyp{ccsubspace}.
Thus, we define it as the closure of the image of \isaterm{S}:
\begin{isabelletop}
lift_definition cblinfun_image
  :: ‹'a[*::complex_normed_vector*] ⇒⇩C⇩L 'b[*::complex_normed_vector*] ⇒ 'a ccsubspace ⇒ 'b ccsubspace›
  is ‹λA S. closure (A ` S)›[* sorry *]
\end{isabelletop}
\idxconst{cblinfun_image}
We prove various basic facts about the interaction between image and other operations such as composition of operators (\isaterm{(A o⇩C⇩L B) *⇩S S =  A *⇩S (B *⇩S S)}),\isalemmafn{cblinfun_compose_image} and intersection of spaces (\isaterm{U *⇩S (inf X Y) = inf (U *⇩S X) (U *⇩S Y)} for isometries~\isaterm{U}).\isalemmafn{isometry_cblinfun_image_inf_distrib}

\paragraph{Adjoints (continued).}\pagelabel{page:bo:adj}
We introduced adjoints already in \autoref{sec:inner-product} (constant \isaconst{cadjoint}) but now we lift the definition to the type \isatyp{'a ⇒⇩C⇩L 'b}:
\begin{isabelletop}
lift_definition adj :: ‹('a::chilbert_space ⇒⇩C⇩L 'b[*::complex_inner*]) ⇒ ('b ⇒⇩C⇩L 'a)›
  is cadjoint[* sorry *]
\end{isabelletop}
\idxconst{adj}
The adjoint \isaterm{adj A} is also written \symbolindexmark{\adjISA}{\isaterm{A*}}.
Since adjoints are only guaranteed to exist over Hilbert spaces, we restrict the domain \isatyp{'a} to be a Hilbert space, so that \isaconst{adj} becomes a total function.

We show important properties of adjoints:
Idempotence,\isalemmafn{double_adj} \isaterm{(A o⇩C⇩L B)* = (B*) o⇩C⇩L (A*)},\isalemmafn{adj_cblinfun_compose} \isaconst{adj} is bounded antilinear,\isalemmafn{bounded_antilinear_adj} \isaterm{norm (A*) = norm A},\isalemmafn{norm_adj} \isaterm{norm (A* o⇩C⇩L A) = (norm A)^2}.\isalemmafn{norm_AadjA}
An operator is called self-adjoint%
\index{self-adjoint}
(\isaterm{selfadjoint A}) when \isaterm{A = A*}.
Then \isaterm{A} is self-adjoint iff \isaterm{ψ ∙⇩C (A *⇩V ψ)} is real for all $\psi$.\isalemmafn{cinner_real_selfadjointI}\fnsep\isalemmafn{cinner_selfadjoint_real}\paragraph{Unitaries, isometries.}%
\index{unitary!(bounded operator)}%
\index{isometry!(bounded operator)}
Using adjoints, we can easily define unitaries and isometries.
$U$ is unitary iff its adjoint is its inverse.
And $U$ is an isometry iff its adjoint is its left-inverse.
\begin{isabelletop}
definition ‹isometry U ⟷ U* o⇩C⇩L U = id_cblinfun›[* for isometry *]
definition ‹unitary U ⟷ U* o⇩C⇩L U = id_cblinfun ∧ U o⇩C⇩L U* = id_cblinfun›[* for unitary *]
\end{isabelletop}
The following alternative characterizations tell us more about the meaning of those unitaries and isometries: An isometry is a norm preserving operator.\isalemmafn{norm_preserving_isometry} Or: $U$ is an isometry if it preserves inner products on a generating set:
\begin{isabelletop}
lemma orthogonal_on_basis_is_isometry:
  assumes ‹ccspan B = ⊤›
  assumes ‹⋀b c. b∈B ⟹ c∈B ⟹ cinner (U *⇩V b) (U *⇩V c) = cinner b c›
  shows ‹isometry U›[* sorry *]
\end{isabelletop}
\idxlemma{orthogonal_on_basis_is_isometry}
And $U$ is unitary iff it is an isometry with dense image:\pagelabel{page:surj_isometry_is_unitary}
\begin{isabelletop}
lemma surj_isometry_is_unitary:
  assumes ‹isometry U› and ‹U *⇩S ⊤ = ⊤› shows ‹unitary U›[* sorry *]
\end{isabelletop}
\idxlemma{surj_isometry_is_unitary}

\paragraph{Sandwiches.}\index{sandwich}%
A common pattern when working with bounded operators is to map \isaterm{B} to \isaterm{A o⇩C⇩L B o⇩C⇩L A*}.
We are not aware of an established name for this, so we simply refer to this as a sandwich.
(When \isaterm{A} is unitary, this mapping is known as a spatial isomorphism\index{spatial isomorphism}\index{isomorphism!spatial}.)
Since for given \isaterm{A}, this is a bounded operator on the space of bounded operators, we define:
\begin{isabelletop}
lift_definition sandwich :: ‹('a::chilbert_space ⇒⇩C⇩L 'b::complex_inner) ⇒
                                            (('a ⇒⇩C⇩L 'a) ⇒⇩C⇩L ('b ⇒⇩C⇩L 'b))›
  is ‹λ(A::'a⇒⇩C⇩L'b) B. A o⇩C⇩L B o⇩C⇩L A*›[* sorry *]
\end{isabelletop}
\idxconst{sandwich}%
Then \isaterm{sandwich A *⇩V B = A o⇩C⇩L B o⇩C⇩L A*}.\isalemmafn{sandwich_apply}
We will use this as a convenient abbreviation in many places throughout the library.

\paragraph{Orthogonal projections (continued).}%
\pagelabel{page:cbo.proj}%
\index{orthogonal projection!(bounded operator)}%
We introduced (orthogonal) projections (a.k.a.~projectors) in \autoref{sec:inner-product}.
We saw that projections onto closed subspaces (of a Hilbert space) always exist and are bounded operators.
For a closed subspace \isaterm{S :: 'a ccsubspace}, we thus define \isaterm{Proj S :: 'a ⇒⇩C⇩L 'a[*::chilbert_space*]} to be the projection onto \isaterm{S}. And we define \isaterm{is_Proj P} to be true iff \isaterm{P} is a projection (onto some set \isaterm{M}):
\begin{isabelletop}
  lift_definition Proj :: ‹'a[*::chilbert_space*] ccsubspace ⇒ 'a ⇒⇩C⇩L 'a› is projection[* sorry *]
  lift_definition is_Proj :: ‹'a[*::complex_normed_vector*] ⇒⇩C⇩L 'a ⇒ bool›
  is ‹λP. ∃M. is_projection_on P M›[* sorry *]
\end{isabelletop}
  \idxconst{Proj}\idxconst{is_Proj}
  (Note that in \isaconst{is_Proj}, we do not explicitly require its range \isaterm{M} to be a closed subspace, but it always is one.\isalemmafn{complex_vector.linear_subspace_image}\fnsep\isalemmafn{Proj_range_closed})

However, now that we have the concept of adjoints, and have restricted our attention to bounded operators, we have a much simpler characterization of projectors, namely self-adjoint idempotent operators:\pagelabel{page:is_Proj_algebraic}
\begin{isabelletop}
lemma is_Proj_algebraic: ‹is_Proj P ⟷ P o⇩C⇩L P = P ∧ P = P*›[* sorry *]
\end{isabelletop}
We can also easily express the one-to-one correspondence between projectors and subspaces: \isaterm{Proj S} is the projector corresponding to a subspace, and given a projector \isaterm{P}, the corresponding subspace is \isaterm{P *⇩S ⊤}. There operations are inverses in the sense that \isaterm{Proj S *⇩S ⊤ = S}\fnsep\isalemmafn{Proj_range} and \isaterm{Proj (P *⇩S ⊤) = P}\fnsep\isalemmafn{Proj_on_own_range} for projectors \isaterm{P}.

We prove various useful laws about projectors, e.g., that their norm is bounded by $1$,\isalemmafn{norm_is_Proj} that \isaterm{1 - P} is the projector onto the orthogonal complement of what \isaterm{P} projects onto,\isalemmafn{Proj_ortho_compl} and more.

\paragraph{Kernels, eigenspaces.}\index{kernel}\index{eigenspace}
The kernel of an operator (i.e., the vectors that are mapped to $0$) is a closed subspace,%
\isalemmafn{kernel_is_closed_csubspace}
so we can define the kernel as a \isatyp{ccsubspace}:
\begin{isabelletop}
lift_definition kernel :: ‹'a[*::complex_normed_vector*] ⇒⇩C⇩L 'b[*::complex_normed_vector*] ⇒ 'a ccsubspace› is ‹λf. f -` {0}›[* sorry *]
\end{isabelletop}
(\isaterm{f -` A} is Isabelle/HOL notation for $f^{-1}(A)$.)
This also allows us to easily define the eigenspace of eigenvalue \isaterm{a} as \isaterm{eigenspace a A = kernel (A - a *⇩C id_cblinfun)}\idxconst{eigenspace}.

One interesting fact about the kernel is that it is the orthogonal complement of the image of the adjoint:
\begin{isabelletop}
lemma kernel_compl_adj_range: ‹kernel a = - (a* *⇩S top)›[* sorry *]
\end{isabelletop}
\idxlemma{kernel_compl_adj_range}%

\paragraph{Partial isometries.}%
\index{partial isometry!(bounded operator)}%
\index{isometry!partial (bounded operator)}
An isometry is a length preserving map on a Hilbert space.
A projector is the identity on a subspace of a Hilbert space.
The adjoint of an isometry is a surjective length-preserving map on a subspace.
What all of these have in common is that they are a length-preserving map on a subspace (and $0$ on its complement).
This is called a partial isometry:
\begin{isabelletop}
definition ‹partial_isometry A
    ⟷ (∀h ∈ space_as_set (- kernel A). norm (A h) = norm h)›[* for partial_isometry *]
\end{isabelletop}
\idxconst{partial_isometry}%
We show that projectors, isometries, and unitaries are partial isometries,\isalemmafn{is_Proj_partial_isometry}\fnsep\isalemmafn{isometry_partial_isometry}\fnsep\isalemmafn{unitary_partial_isometry} that partial isometries are closed under adjoints,\isalemmafn{partial_isometry_adj} that they have norm $1$ (with the exception of the zero-operator),\isalemmafn{norm_partial_isometry} and that \isaterm{A* o⇩C⇩L A} is the projector onto the initial space\footnote{I.e., the complement of the kernel.} of a partial isometry~\isaterm{A}.\isalemmafn{partial_isometry_adj_a_o_a}

\paragraph{Isomorphisms and inverses.}%
\index{isomorphism!(bounded operator)}%
\index{inverse!(bounded operator)}%
\index{left-inverse!(bounded operator)}
We call a bounded operator \isaterm{A} invertible%
\index{invertible!(bounded operator)} if it has a left inverse that is a bounded operator, too:
\begin{isabelletop}
definition ‹invertible_cblinfun A ⟷ (∃B. B o⇩C⇩L A = id_cblinfun)›[* for invertible_cblinfun *]
\end{isabelletop}
\idxconst{invertible_cblinfun}%
If it has a two-sided inverse, we call it an isomorphism (\isaterm{iso_cblinfun A}\idxconst{iso_cblinfun}).
We define the left-inverse of an operator as an arbitrary left-inverse (if it exists).
\begin{isabelletop}
definition ‹cblinfun_inv A = (if invertible_cblinfun A then
                              (SOME B. B o⇩C⇩L A = id_cblinfun) else 0)›[* for cblinfun_inv *]
\end{isabelletop}
\idxconst{cblinfun_inv}%
Obviously, for \isaterm{invertible_cblinfun A}, \isaterm{cblinfun_inv A} is a left-inverse,\isalemmafn{cblinfun_inv_left}
but if \isaterm{iso_cblinfun A}, then \isaterm{cblinfun_inv A} is also a \emph{right}-inverse.\isalemmafn{cblinfun_inv_right}
Both-sided inverses are unique.\isalemmafn{cblinfun_inv_uniq}
Applying an invertible bounded operator commutes with taking an infinite sum: \isaterm{(∑⇩∞x∈S. A *⇩V g x) = A *⇩V (∑⇩∞x∈S. g x)}.\isalemmafn{infsum_cblinfun_apply_invertible}

\paragraph{Extension of bounded operators.}\pagelabel{page:cblinfun_extension}%
\index{extension of bounded operators}%
\index{bounded operators!extension of}%
Often, we want to construct an operator by specifying it on a subset (e.g., on a basis) and then define the operator by linear extension.
E.g., if \isaterm{{ket 0, ket 1}} is a basis of the space,\footnote{%
  The constant \isaterm{ket} will be formally introduced later.} then we might define \isaterm{X} by specifying \isaterm{X *⇩V ket 0 = ket 1} and \isaterm{X *⇩V ket 1 = ket 0} and say ``\isaterm{X} is defined everywhere else by linearity''.
For linear (not necessarily bounded) operators, Isabelle/HOL already has an existing constant \isaconstidx{cconstruct} for this purpose that we will not describe further here (not a contribution of our library).
In the case of bounded operators, the situation is more complex, however.
An operator specified on a basis may have a linear extension but that one might not be bounded.
E.g., if \isaterm{ket i} (with \isaterm{i::nat}) form an orthonormal basis, then there is a linear operator with \isaterm{A *⇩V ket i = ket 0} for all \isaterm{i}, but it is not bounded.\isalemmafn{bounded_extension_counterexample_1}\fnsep\isalemmafn{bounded_extension_counterexample_2}
To cover this use case, we introduce a constant \isaconstidx{cblinfun_extension}:
If \isaterm{φ} is a function defined on \isaterm{S}, then \isaterm{cblinfun_extension S φ} is some bounded operator that coincides with \isaterm{φ} on \isaterm{S} -- if such an operator exists.
\begin{isabelletop}
definition ‹cblinfun_extension S φ = (SOME B. ∀x∈S. B *⇩V x = φ x)›[* for cblinfun_extension *]
\end{isabelletop}

We give various criteria for when this operator exists.
To be able to state them compactly, we introduce the constant \isaconstidx{cblinfun_extension_exists}:
\begin{isabelletop}
definition ‹cblinfun_extension_exists S φ = (∃B. ∀x∈S. B *⇩V x = φ x)›[* for cblinfun_extension_exists *]
\end{isabelletop}
If \isaterm{cblinfun_extension_exists S φ}, we have \isaterm{cblinfun_extension S φ *⇩V v = φ v} for all \isaterm{v ∈ S}.\isalemmafn{cblinfun_extension_apply} The extension is unique if the span of \isaterm{S} is dense.\isalemmafn{separating_set_bounded_clinear_dense}

We then have:
\begin{itemize}
\item The so-called BLT theorem:\index{BLT theorem}
  If \isaterm{φ} is bounded-linear on a dense subspace \isaterm{S}, and the codomain of \isaterm{φ} is a Banach space, then \isaterm{φ} can be extended to the whole space:
  \begin{isabelletop}
lemma cblinfun_extension_exists_bounded_dense:
  fixes f :: ‹'a::complex_normed_vector ⇒ 'b::cbanach›
  assumes ‹csubspace S› and ‹closure S = UNIV›
  assumes ‹⋀x y. x ∈ S ⟹ y ∈ S ⟹ f (x + y) = f x + f y›
  assumes ‹⋀c x y. x ∈ S ⟹ f (c *⇩C x) = c *⇩C f x›
  assumes ‹⋀x. x ∈ S ⟹ norm (f x) ≤ B * norm x›
  shows ‹cblinfun_extension_exists S f›[* sorry *]
\end{isabelletop}
\idxlemma{cblinfun_extension_exists_bounded_dense}%
  This is probably one of the most important tools for constructing bounded operators.
  Also in this case, \isaterm{norm (cblinfun_extension S f) ≤ B}.\isalemmafn{cblinfun_extension_norm_bounded_dense}
\item The BLT theorem still requires us to define \isaterm{φ} on a dense subspace and to show that it is bounded-linear there.
  On inner product spaces, we have the following simpler corollary:
  If \isaterm{S} is an orthogonal basis, and \isaterm{φ} is a bounded function from \isaterm{S} into a Hilbert space that maps different basis elements to orthogonal vectors, then \isaterm{cblinfun_extension_exists S f}.\isalemmafn{cblinfun_extension_exists_ortho}
\item In the finite-dimensional case, things are even simpler: We do not require orthogonality. If \isaterm{S} is a basis, then \isaterm{cblinfun_extension_exists S f}.\isalemmafn{cblinfun_extension_exists_finite_dim}
\item One further limitation of the BLT theorem is that it requires that \isaterm{S} is dense.
  On Hilbert spaces, we can omit this condition.\isalemmafn{cblinfun_extension_exists_hilbert}
\item There are a few more cases but we omit them here for space reasons.
  Searching for theorems involving \isaconst{cblinfun_extension_exists} reveals those lemmas.
\end{itemize}

Note that over the type \isatyp{ell2} (introduced in \autoref{sec:ell2} below), we introduce further constants \isaconst{explicit_cblinfun} (\autopageref{page:explicit_cblinfun}) and \isaconst{classical_operator} (\autopageref{page:classical_operator}) for defining bounded operators that can be easier to use in certain cases.

\paragraph{Product spaces (continued).}\index{product spaces}%
\index{product space!(of inner product / Hilbert space; ctd.)}
On \autopageref{page:product-spaces}, we showed that the product of two normed vector spaces is a normed vector space.
We now introduce two important bounded operators into product spaces, mapping $\psi$ to $(\psi,0)$ and $(0,\psi)$, respectively (canonical embeddings):
\begin{isabelletop}
lift_definition cblinfun_left :: ‹'a[*::complex_normed_vector*] ⇒⇩C⇩L ('a×'b[*::complex_normed_vector*])› is ‹(λx. (x,0))›[* sorry *]
\end{isabelletop}
\idxconst{cblinfun_left}%
and analogously for \isaconstidx{cblinfun_right}. Since they are isometries,\isalemmafn{isometry_cblinfun_left} their inverses are \isaterm{cblinfun_left*} and \isaterm{cblinfun_right*}.

Why are these constants useful? In non-formalized mathematics, one will often encounter matrices of operators such as, e.g., {\tiny$\begin{pmatrix} 0 & A \\ B & 0 \end{pmatrix}$} for bounded operators $A,B$.
Constructing this operator from $A$ and $B$ by hand is tedious, one needs to prove the boundedness directly and then use \isaconst{cblinfun_extension} to construct it.
With the constants introduced here, we can write this matrix as \isaterm{cblinfun_left o⇩C⇩L A o⇩C⇩L cblinfun_right* + cblinfun_right o⇩C⇩L B o⇩C⇩L cblinfun_left*}, and all properties of bounded operators follow from the type system.
Working with such matrices is then made easy by having various simplification rules, e.g., when multiplying two of them, many terms cancel out due to rules like \isaterm{cblinfun_right* o⇩C⇩L cblinfun_left = 0}.\isalemmafn{cblinfun_right_left_ortho}
We stress that this approach only works for small matrices (the size of the nested \isaconst{cblinfun_left}/\isaconst{cblinfun_right} calls grows linearly with the matrix size).
For constructions involving large or unbounded matrices, one will have to take the route through \isaconst{cblinfun_extension}.

Furthermore, the Cartesian product of two closed subspaces is a closed subspace, so we can define the product of \isatyp{ccsubspace}'s:
\begin{isabelletop}
lift_definition ccsubspace_Times
  :: ‹'a[*::complex_normed_vector*] ccsubspace ⇒ 'b[*::complex_normed_vector*] ccsubspace ⇒ ('a×'b) ccsubspace›
  is ‹λS T. S × T›[* sorry *]
\end{isabelletop}
\idxconst{ccsubspace_Times}%
We add various lemmas about bases of product spaces and product subspace.\isalemmafn{is_ortho_set_prod}\fnsep\isalemmafn{is_onb_prod}\fnsep\isalemmafn{ccsubspace_Times_ccspan}

\paragraph{Positive operators and the Loewner order (definition).}%
\pagelabel{page:def-pos-ops}%
\index{Loewner order!(definition)}\index{order!Loewner (definition)}%
\index{positive operator!(definition)}\index{operator!positive (definition)}
Readers not interested in the technical challenges encountered when defining the Loewner order can skip to the next subsection.
There we cover the properties of the Loewner order.

A square\footnote{%
  A \emph{square operator}\index{square operator}\index{operator!square} is one with same domain and codomain, i.e., of type \isatyp{'a ⇒⇩C⇩L 'a} as opposed to \isatyp{'a ⇒⇩C⇩L 'b}.}
bounded operator \isaterm{A} is called positive iff \isaterm{ψ ∙⇩C (A *⇩V ψ) ≥ 0} for all \isaterm{ψ}.
Besides being a very important concept in its own right, it also gives rise to a natural ordering of operators, the Loewner order:
\isaterm{A ≤ B} iff \isaterm{B - A} is positive.
At the first glance, it would seem trivial to cast this into definitions in Isabelle/HOL.
And indeed, it is easy to define a constant \isaterm{pos_op :: ('a ⇒⇩C⇩L 'a) ⇒ bool}.
But when defining the Loewner order (based on \isaterm{pos_op}), we would like to be able to write \isaterm{A ≤ B} (use the overloaded constant \isaconst{less_eq}) and not have to write, e.g., \isatermpre{consts less_eq_op::'a}{less_eq_op A B}.
Besides notational inconvenience, which could be overcome by adding some extra syntax, the problem is that without overloading \isaconst{less_eq}, we cannot tell Isabelle that the Loewner order makes the type \isatyp{cblinfun} into an ordered type (type class \isaclass{order}) and thus miss out on all the advantages of type classes and the possibility to reuse existing results about ordered types. An extra benefit of defining $\leq$ as the Loewner order is that we can then write ``\isaterm{A} is positive'' simply as ``\isaterm{A ≥ 0}'', obviating the need for an extra constant \isaterm{pos_op}.

Unfortunately, we cannot write a definition of the following form: \isaterm{A ≤ B ⟷ (∀ψ. ψ ∙⇩C (A *⇩V ψ) ≤ ψ ∙⇩C (B *⇩V ψ))} for \isaterm{A}, \isaterm{B :: 'a ⇒⇩C⇩L 'a}.
This is because the type-class instantiation mechanism of Isabelle can only be applied to types such as \isatyp{'a ⇒⇩C⇩L 'b} where we allow \isatyp{'a} and \isatyp{'b} to be different.\footnote{%
  Strictly speaking, defining the constant \isaconst{less_eq} is possible but the subgoal created by the \isatop{instance[*int::finite sorry*]} command would be impossible to prove because they will refer to the constant with a more general type.}
Restricting a type-class instantiation to square operators is not possible.
But  our definition of $\leq$ is not well-typed for \isaterm{A}, \isaterm{B :: 'a ⇒⇩C⇩L 'b}.

Our approach is to give a definition of $\leq$ on bounded operators that is well-defined (if not particularly mathematically meaningful) on non-square operators, and has the desired definition as a special case for square operators.
The first step is to define a ``heterogeneous identity'',%
\index{heterogeneous identity}%
\index{identity!heterogeneous}
i.e., a function that is the identity if domain and codomain are the same type, and not further specified otherwise.
(This will in the end allow us to identify different domains and codomains.)
Isabelle allows us to do this with the following commands:
\begin{isabelletop}
consts heterogenous_identity :: ‹'a ⇒ 'b›
overloading heterogenous_identity_id ≡ ‹heterogenous_identity :: 'a ⇒ 'a›
begin definition ‹heterogenous_identity_id = id› end
\end{isabelletop}
\idxconst{heterogenous_identity}%
Now \isaterm{heterogenous_identity = id} holds but only when \isaterm{heterogenous_identity} is restricted to the type \isatyp{'a ⇒ 'a}. 
Next we need a heterogeneous equality, but as a bounded linear functions.
(\isaconst{heterogenous_identity} will not, in general, be bounded linear when \isatyp{'a}$\neq$\isatyp{'b}.)
We define this by placing a wrapper around \isaconst{heterogenous_identity}, making it to be of type \isatyp{'a ⇒⇩C⇩L 'b}:
\begin{isabelletop}
lift_definition heterogenous_cblinfun_id :: ‹'a[*::complex_normed_vector*] ⇒⇩C⇩L 'b[*::complex_normed_vector*]› is
  ‹if bounded_clinear (heterogenous_identity :: 'a ⇒ 'b)
   then heterogenous_identity else (λ_. 0)›[* sorry *]
\end{isabelletop}
\idxconst{heterogenous_cblinfun_id}%
The fallback \isaterm{(λ_. 0)} guarantees that \isaconst{heterogenous_cblinfun_id} can always be typed as a \isatyp{cblinfun}, even if \isaconst{heterogenous_identity} is not bounded linear.
Since for \isatyp{'a}=\isatyp{'b}, we have \isaterm{heterogenous_identity = id} and \isaterm{id} is bounded linear, we get \isaterm{heterogenous_cblinfun_id = id_cblinfun} in this case.\isalemmafn{heterogenous_cblinfun_id_def'}.
(And we do not care about other cases.)
When using \isaconst{heterogenous_cblinfun_id} in the definition of the Loewner order, we will need to make a case distinction between the cases \isatyp{'a}=\isatyp{'b} and \isatyp{'a}$\neq$\isatyp{'b}.
Unfortunately, such a case distinction cannot be expressed in Isabelle/HOL.
Instead, we roughly approximate \isatyp{'a}=\isatyp{'b} by ``\isatyp{'a} and \isatyp{'b} are unitarily isomorphic (and \isaconst{heterogenous_cblinfun_id} is the isomorphism)''.
We write this as \isaterm{heterogenous_same_type_cblinfun TYPE('a[*::chilbert_space*]) TYPE('b[*::chilbert_space*])}, defined as follows:
\begin{isabelletop}
definition ‹heterogenous_same_type_cblinfun (x::'a[*::chilbert_space*] itself) (y::'b[*::chilbert_space*] itself)
  ⟷ unitary (heterogenous_cblinfun_id :: 'a ⇒⇩C⇩L 'b)
    ∧ unitary (heterogenous_cblinfun_id :: 'b ⇒⇩C⇩L 'a)›[* for heterogenous_same_type_cblinfun *]
\end{isabelletop}
\idxconst{heterogenous_same_type_cblinfun}%
Note that then \isaterm{heterogenous_same_type_cblinfun TYPE('a[*::chilbert_space*]) TYPE('a)} holds.\isalemmafn{heterogenous_same_type_cblinfun}

We can now state the definition of the Loewner order:
\begin{isabelletop}
[* (* To avoid `Bad head of lhs: existing constant "(≤)"` message: *)
no_notation less_eq  (‹(‹notation=‹infix ≤››_/ ≤ _)›  [51, 51] 50)
syntax "_less_eq" :: ‹'a ⇒ 'a ⇒ 'a› ("(_/ ≤ _)"  [51, 51] 50)
parse_translation ‹[("_less_eq", fn ctxt => fn [t,u] => Free("dummy", dummyT) $ t $ u)]›
*]definition ‹A ≤ B ⟷
  (if heterogenous_same_type_cblinfun TYPE('a[*::chilbert_space*]) TYPE('b[*::chilbert_space*]) then
    ∀ψ::'b. ψ ∙⇩C ((B-A) *⇩V heterogenous_cblinfun_id *⇩V ψ) ≥ 0 else (A=B))›
\end{isabelletop}
Note that when \isatyp{'a}=\isatyp{'b}, the if-condition is true and \isaconst{heterogenous_cblinfun_id} is the identity, so this simplifies to:
\begin{isabelletop}
lemma less_eq_cblinfun_def:
  ‹A ≤ B ⟷ (∀ψ. ψ ∙⇩C (A *⇩V ψ) ≤ ψ ∙⇩C (B *⇩V ψ))›[* sorry *]
\end{isabelletop}
\idxlemma{less_eq_cblinfun_def}
So in this case, we have indeed defined the Loewner order.

However, note that $\leq$ is now also defined (if somewhat arbitrarily) on \isatyp{'a ⇒⇩C⇩L 'b} for \isatyp{'a}$\neq$\isatyp{'b}.
Either \isaterm{heterogenous_same_type_cblinfun TYPE('a[*::chilbert_space*]) TYPE('a)} holds, then $\leq$ is the Loewner order up to a (somewhat arbitrary) identification of the isomorphic types \isatyp{'a} and \isatyp{'b}.
Or it doesn't hold, then $\leq$ is the equality.
In both cases, we can still derive that $\leq$ is, e.g., a partial order (type class \isaclass{order}) since both the Loewner order and equality are partial orders.
And similarly, \isatyp{cblinfun} is an instance of any type class that both the Loewner order and equality satisfy.
Since equality is quite well-behaved, we expect this to work for most type classes that we would expect the Loewner order to satisfy.Note that due to the use of the constant \isaconst{unitary} in the definition, all the above only works when \isatyp{'a :: chilbert_space}, not for arbitrary inner product spaces.
It might be possible to refine the approach above to get the definition of the Loewner order to work without assuming Hilbert spaces but we have not explored this.

\paragraph{Positive operators and the Loewner order (properties).}
\index{Loewner order!(properties)}\index{order!Loewner (properties)}%
\index{positive operator!(properties)}\index{operator!positive (properties)}
A square bounded operator \isaterm{A} is called positive iff \isaterm{ψ ∙⇩C (A *⇩V ψ) ≥ 0} for all \isaterm{ψ}.
Besides being a very important concept in its own right, it also gives rise to a natural ordering of operators, the Loewner order:
\isaterm{A ≤ B} iff \isaterm{B - A} is positive.
We thus define the Loewner order as \isaterm{A ≤ B ⟷ (∀ψ. ψ ∙⇩C (A *⇩V ψ) ≤ ψ ∙⇩C (B *⇩V ψ))}.
(See the previous subsection for Isabelle-specific subtleties involved here, and an explanation why we only define this for operators over \textit{Hilbert spaces}.)
We can then write \isaterm{A ≥ 0} to denote that \isaterm{A} is positive.

The Loewner order is, as the name says, a partial order. It makes \isatyp{'a ⇒⇩C⇩L 'b} into an ordered complex vector space:
\begin{isabelletop}
instance cblinfun :: (chilbert_space, chilbert_space) ordered_complex_vector[* sorry *]
\end{isabelletop}
(The type class \isaclassidx{ordered_complex_vector} inherits from both \isaclass{complex_vector} and \isaclass{order} and ensures that the order $\leq$ is compatible with the vector space operations in natural ways.)

We can finally show various properties of positive operators, e.g., the identity is positive,\isalemmafn{positive_id_cblinfun} positive operators are self-adjoint,\isalemmafn{positive_selfadjointI} \isaterm{b* o⇩C⇩L b ≥ 0} (even when \isaterm{b} is not a square operator).\isalemmafn{positive_cblinfun_squareI}
Also, when restricted to projectors (which stand in one-to-one correspondence with closed subspaces), the Loewner order coincides with the natural ordering of subspaces:
\begin{isabelletop}
lemma Proj_mono: ‹Proj S ≤ Proj T ⟷ S ≤ T›[* sorry *]
\end{isabelletop}
\idxlemma{Proj_mono}%
Furthermore, operators between one-dimensional spaces are canonically isomorphic to the complex numbers (via \isaconst{one_dim_iso}), and in that case the Loewner order coincides with natural ordering of the complex numbers:%
\index{complex numbers!ordering}\index{ordering of the complex numbers}%
\footnote{The ordering on complex numbers, from the theory \isathy[HOL-Library]{Complex_Order} in the Isabelle distribution, is defined by \isaterm{x ≤ y} iff \isaterm{Re x ≤ Re y} and \isaterm{Im x = Im y}.}
\begin{isabelletop}
lemma one_dim_loewner_order:
  ‹A ≥ B ⟷ one_dim_iso A ≥ (one_dim_iso B :: complex)›[* for A B :: ‹'a ⇒⇩C⇩L 'a::{chilbert_space, one_dim}› sorry*]
\end{isabelletop}
\idxlemma{one_dim_loewner_order}

\paragraph{Vectors as operators.}%
\pagelabel{page:vectors-as-ops}%
\index{vector!(as bounded operator)}
A vector \isaterm{ψ} in $\setC^n$ (i.e., an $n$-dimensional column of complex numbers) is usually identified with an $n\times 1$-matrix.
For example, the inner product \isaterm{ψ ∙⇩C φ} is often written simply as $\isaterm{ψ}^*\cdot \isaterm{φ}$, where taking the adjoint $\isaterm{ψ}^*$ of \isaterm{ψ} is meaningful because we see \isaterm{ψ} as a matrix.

In the general (non-matrix) case, the anologue is to identify \isaterm{ψ :: 'a} with the operator of type \isatyp{complex ⇒⇩C⇩L 'a} mapping \isaterm{c::complex} to \isaterm{c *⇩C ψ}.
Yet, we need to generalize this even further:
In some situations, an operator of type \isatyp{complex ⇒⇩C⇩L 'a} is most natural. In others, \isatyp{unit ell2 ⇒⇩C⇩L 'a} (where \isatyp{unit ell2} means $\setC^1$, see~\autoref{sec:ell2}) is more natural.
Fortunately, all one-dimensional spaces (typeclass \isaclass{one_dim}, see \autoref{sec:one-dim}) are canonically isomorphic to \isatyp{complex} (via \isaconst{one_dim_iso}), so we simply make the definition generic, and allow any one-dimensional space instead of \isatyp{complex}:
\begin{isabelletop}
lift_definition vector_to_cblinfun :: ‹'a[*::complex_normed_vector*] ⇒ 'b::one_dim ⇒⇩C⇩L 'a› is
  ‹λψ φ. one_dim_iso φ *⇩C ψ›[* sorry *]
\end{isabelletop}
\idxconst{vector_to_cblinfun}%
Then \isaterm{vector_to_cblinfun ψ} is the same as \isaterm{ψ}, except it can be given type \isatyp{'b ⇒⇩C⇩L 'a}, and \isaprop{vector_to_cblinfun ψ *⇩V φ = one_dim_iso ψ *⇩C φ}.\isalemmafn{vector_to_cblinfun_apply}
The condition is that \isatyp{'b} is a one-dimensional space (including \isatyp{complex} and \isatyp{unit ell2}).

Note that \isaterm{vector_to_cblinfun ψ} is a bounded operator due to its type, and an isometry if \isaterm{ψ} is a unit vector.\isalemmafn{isometry_vector_to_cblinfun}
The map \isaconst{vector_to_cblinfun} (as a function of \isaterm{ψ}) is also bounded linear\isalemmafn{bounded_clinear_vector_to_cblinfun} and norm-preserving,\isalemmafn{norm_vector_to_cblinfun} and its image is \isaterm{ccspan {ψ}}.\isalemmafn{image_vector_to_cblinfun}

We also prove various compatibility laws with \isaterm{(∙⇩C)}, \isaterm{(*⇩V)}, \isaterm{(o⇩C⇩L)}, etc., which allow us to simplify terms such as \isaterm{(vector_to_cblinfun a)* *⇩V b} or \isaterm{(vector_to_cblinfun a)* o⇩C⇩L vector_to_cblinfun b} directly to the inner product \isaterm{a ∙⇩C b}.

\paragraph{Rank-1 operators.}%
\index{rank-1 operator}%
\index{operator!rank-1}%
\label{page:rank1}
A simple but important class of operators are rank-1 operators, operators whose image is a one-dimensional subspace. We deviate slightly from the usual definition, and define a rank-1 operator as one that has a one- or zero-dimensional image, i.e., the span of a possibly zero vector:
\begin{isabelletop}
  definition ‹rank1 A ⟷ (∃ψ. A *⇩S ⊤ = ccspan {ψ})›[* for rank1 *]
\end{isabelletop}
\idxconst{rank1}
(A rank-1 operator in the usual sense is then one with \isaterm{rank1 A ∧ A≠0}.
Our nonstandard definition leads to less case distinctions down the line.)

The importance of rank-1 operators arises from two facts:
The important compact operators%
\index{compact operator}%
\index{operator!compact}
are the closed span of the rank-1 operators \cite[Thm.~II.4.4]{conway2013course}.
(We do not prove that fact in this formalization.)
And the rank-1 operators correspond, on Hilbert spaces, to the very natural operators of the form $\isaterm{ψ}\ \isaterm{φ}^*$ for vectors \isaterm{ψ}, \isaterm{φ}, mapping \isaterm{φ} onto \isaterm{ψ} (assuming unit \isaterm{φ}).
In quantum mechanics, these are usually written $\ket\psi\bra\phi$ which is visually reminiscent of a butterfly;\index{butterfly} we adopt this name here:
\begin{isabelletop}
definition
  ‹butterfly ψ φ = vector_to_cblinfun ψ o⇩C⇩L (vector_to_cblinfun φ[* :: complex ⇒⇩C⇩L _*])*›[* for butterfly *]
\end{isabelletop}
\idxconst{butterfly}
(Note that we need to use \isaconst{vector_to_cblinfun} from the previous section here to write $\isaterm{ψ}\ \isaterm{φ}^*$.)
We show that these ``butterflies'' are exactly the rank-1 operators.\isalemmafn{rank1_iff_butterfly}

Rank-1 operators are closed under composition with bounded operators,\isalemmafn{rank1_compose_left}\fnsep\isalemmafn{rank1_compose_right} and scaling,\isalemmafn{rank1_scaleC} and under adjoints.\isalemmafn{rank1_adj}

Butterflies interact well with \isaterm{(∙⇩C)}, \isaterm{(*⇩V)}, \isaterm{(o⇩C⇩L)}, and adjoints; we prove various simplification rules for this.
The norm of \isaterm{butterfly ψ φ} is the product of the norms of \isaterm{ψ}, \isaterm{φ},\isalemmafn{norm_butterfly} and \isaterm{butterfly} is bounded sesquilinear.\isalemmafn{bounded_sesquilinear_butterfly}

One important use of butterflies is that they give us a simple way to write (and reason about) projectors: \isaterm{butterfly ψ ψ} is the projector onto \isaterm{ψ} (assuming a unit vector~\isaterm{ψ}),\isalemmafn{butterfly_eq_proj} and the projector onto the space spanned by finitely many orthogonal vectors can be represented as a sum of such ``self-butterflies''.\isalemmafn{sum_butterfly_is_Proj}

Finally, butterflies are quite useful for reasoning about operators on finite dimensional spaces:
If \isaterm{A} and \isaterm{B} are orthonormal sets of vectors, then \isaterm{butterfly a b} with \isaterm{a ∈ A}, \isaterm{b ∈ B} form an orthonormal set, too.\isalemmafn{cindependent_butterfly}
And if \isaterm{A} and \isaterm{B} span (not closed span!) the whole space, so do the \isaterm{butterfly a b}.\isalemmafn{cspan_butterfly_UNIV}
As an immediate consequence, linear functions on the space of bounded operators are equal iff they agree on all those \isaterm{butterfly a b} (assuming finite dimensional Hilbert spaces).\isalemmafn{clinear_eq_butterflyI}\fnsep\footnote{%
  Note that in the infinite-dimensional case the analogous result (with closed spans) does not hold, since the closed span of the rank-1 operators are only the compact operators.} Also in finite-dimensional spaces, the \isaterm{butterfly a a} sum to \isaconst{id_cblinfun}.\isalemmafn{butterflies_sum_id_finite}

\paragraph{Banach--Steinhaus theorem.}
The \emph{Banach--Steinhaus theorem}\index{Banach--Steinhaus theorem}, also known as the \emph{principle of uniform boundedness}\index{principle of uniform boundedness}\index{uniform boundedness!principle of} is a central theorem of functional analysis.
It states that if a family of bounded operators is bounded pointwise (i.e., $\norm{F_nx}$ is bounded for every unit vector $x$), then the family is bounded uniformly (i.e., $\norm{F_n}$ is bounded).
In Isabelle:
\begin{isabelletop}
theorem cbanach_steinhaus:
  fixes F :: ‹'c ⇒ 'a::cbanach ⇒⇩C⇩L 'b[*::complex_normed_vector*]›
  assumes ‹⋀x. ∃M. ∀n.  norm ((F n) *⇩V x) ≤ M›
  shows  ‹∃M. ∀n. norm (F n) ≤ M›[* sorry *]
\end{isabelletop}
\idxlemma{cbanach_steinhaus}%
This theorem was already shown in \cite{Banach_Steinhaus-AFP} for \emph{real} bounded operators. We can lift it to the complex case by using the fact that every complex bounded operator is also a real bounded operator (using Isabelle's ``transfer'' mechanism \cite{transfer}).

\paragraph{Riesz representation theorem (ctd.).}
\index{Riesz representation theorem!(bounded operators)}
We already introduced the \emph{Riesz representation\index{Riesz representation} theorem} in \autoref{sec:inner-product}, \autopageref{page:riesz-rep}.
Using the type \isatyp{cblinfun}, we can represent it more compactly. For example, existence becomes simply: \isaprop{∃t. ∀x.  f *⇩V x = (t ∙⇩C x)}\isalemmafn{riesz_representation_cblinfun_existence}.
(Here \isaterm{f :: _ ⇒ complex} by type inference, i.e., \isaterm{f} is a bounded linear functional.)
In addition, we show that the norm of \isaterm{t} equals that of~\isaterm{f}.\isalemmafn{riesz_representation_cblinfun_norm}
Due to the uniqueness of \isaterm{t}, we can define the function that maps \isaterm{f} to \isaterm{t}:
\begin{isabelletop}
definition the_riesz_rep :: ‹('a[*::chilbert_space*] ⇒⇩C⇩L complex) ⇒ 'a› where
  ‹the_riesz_rep f = (SOME t. ∀x. f *⇩V x = t ∙⇩C x)›
\end{isabelletop}
\idxconst{the_riesz_rep}%
Then \isaterm{the_riesz_rep f ∙⇩C x = f *⇩V x},\isalemmafn{the_riesz_rep}, and \isaconst{the_riesz_rep} is bounded antilinear.\isalemmafn{bounded_antilinear_the_riesz_rep}

Finally, if we have a family \isaterm{p} of bounded linear functions, and apply the Riesz-representation theorem pointwise, we get a family of vectors.
If \isaterm{p} (see as a two-argument function) is bounded sesquilinear, that family of vectors (seen as a function) turns out to be a bounded operator.
We capture this by an additional definition \isaterm{the_riesz_rep_sesqui :: ('a[*::complex_normed_vector*] ⇒ 'b[*::chilbert_space*] ⇒ complex) ⇒ ('a ⇒⇩C⇩L 'b)}\idxconst{the_riesz_rep_sesqui} with the defining property \isaterm{(the_riesz_rep_sesqui p *⇩V x) ∙⇩C y = p x y} whenever \isaterm{p} is sesquilinear.\isalemmafn{the_riesz_rep_sesqui_apply}
(Note that the type of \isaconst{the_riesz_rep_sesqui} expresses that the resulting family is a bounded operator.)
In other words, this captures the fact that any (sesquilinear) inner product can be written as $x^*Ay$ for suitable $A$.

\paragraph{The bidual.}
The \emph{dual}\index{dual} of a normed vector space \isatyp{'a} is the set of bounded linear functionals on \isatyp{'a}.
Or, using the types defined in our development, the dual is simply \isatyp{'a ⇒⇩C⇩L complex}.
The dual of the dual is called the \emph{bidual}\index{bidual}: \isatyp{('a ⇒⇩C⇩L complex) ⇒⇩C⇩L complex}.
In general, there is no canonical way to interpret \isaterm{ψ :: 'a} as an element of the dual,%
\footnote{It is canonical to identify \isaterm{ψ} with $\isaterm{φ} \mapsto \isaterm{ψ ∙⇩C φ}$ but that only works on inner product spaces.}
but there is a canonical embedding of \isatyp{'a} into its bidual, namely $\psi$ is identified with the function $f\mapsto f\psi$. This identification is, in turn, a bounded linear map, so we can define it as follows:
\begin{isabelletop}
lift_definition bidual_embedding :: ‹'a[*::complex_normed_vector*] ⇒⇩C⇩L (('a ⇒⇩C⇩L complex) ⇒⇩C⇩L complex)›
  is ‹λx f. f *⇩V x›[* sorry *]
\end{isabelletop}
(Note how all the arrows in the type are \isatyp{[*_*]⇒⇩C⇩L[*_*]}, i.e., all maps involved in this construction are guaranteed by their type to be bounded linear.)
This is indeed an embedding in the sense that \isaconst{bidual_embedding} is an isometry.\isalemmafn{isometry_bidual_embedding}
In general it is not surjective. (Spaces where it is surjective, i.e., that are isometrically isomorphic to their bidual, are called \emph{reflexive}\index{reflexive} Banach spaces.)
However, if \isatyp{'a} is a Hilbert space, then it is a consequence of the Riesz representation that \isaconst{bidual_embedding} is also surjective.\isalemmafn{bidual_embedding_surj}

\paragraph{Bijections between orthogonal bases.}\pagelabel{page:bo:onb}%
\index{orthogonal basis!bijection between}
We already saw that every Hilbert space has an orthonormal basis (ONB), see \autoref{sec:inner-product}, \autopageref{page:inner:onb}.
In general there is not just one ONB.
However, all orthogonal bases of a given Hilbert space have the same cardinality, i.e., there is a bijection between them:
\begin{isabelletop}
lemma all_ortho_bases_same_card:
  assumes ‹is_ortho_set E› and ‹is_ortho_set F›
      and ‹ccspan E = ⊤› and ‹ccspan F = ⊤›
  shows ‹∃f. bij_betw f E F›[* sorry *]
\end{isabelletop}
\idxlemma{all_ortho_bases_same_card}
(Note that logically, this lemma should be in the theory \isathy{Complex_Inner_Product} since it does not refer to anything involving \isatyp{cblinfun}. But its proof depends on some developments in the current theory, so it is placed here.)
For convenience, we define a constant \isaterm{bij_between_bases E F} that is an arbitrary bijection between \isaterm{E} and \isaterm{F}, if both are orthogonal bases of the same space.\isalemmafn{bij_between_bases_bij}
If \isaterm{E}, \isaterm{F} are even ONBs, we extend this by defining a \emph{unitary} operator \isaterm{unitary_between E F} on the whole space that maps \isaterm{E} to \isaterm{F}.\isalemmafn{unitary_between_unitary}\fnsep\isalemmafn{unitary_between_apply}.

\paragraph{Notation control.}\index{notation}
The library introduces some infix notation for the most common types and constants, namely \isatyp{[*_*]⇒⇩C⇩L[*_*]} for the type \isatyp{cblinfun}, \isaterm{[*x*]o⇩C⇩L[*y*]} for the composition \isaconst{cblinfun_compose} of bounded operators, \isaterm{[*x*]*⇩V[*y*]} for the application \isaconst{cblinfun_apply}, \isaterm{[*x*]*⇩S[*y*]} for the image \isaconst{cblinfun_image} of a subspace, \isaterm{[*x*]*} for the adjoint.
And (not specific to bounded operators), \isaterm{[*x*]∙⇩C[*y*]} for the inner product \isaconst{cinner}.

In order to avoid pollution of the global syntax, we make it possible to activate and deactivate the bounded-operator-related syntax using the commands \isatop{unbundle cblinfun_syntax}\idxbundle{cblinfun_syntax} and \isatop{unbundle no cblinfun_syntax}. And, analogous with \isathy[HOL-Analysis]{Inner_Product} from \isasession{HOL-Analysis}, \isatop{unbundle cinner_syntax}\idxbundle{cinner_syntax} and \isatop{unbundle no cinner_syntax} enable and disable \isaterm{[*x*]∙⇩C[*y*]}.


\section{Square summable functions (\texorpdfstring{$\ell_2$}{ℓ₂})}%
\index{square summable functions ($\ell_2$)}%
\index{summable functions!square- ($\ell_2$)}%
\label{sec:ell2}

The archetypal Hilbert space is \symbolindexmark\elltwo{$\elltwo(X)$}, the space of all square-summable functions, i.e., functions $f:X\to\setC$ such that $\sum_{x\in X}\abs{f(x)}^2<\infty$.
This Hilbert space has the orthonormal basis $\{e_x\}_{x\in X}$ where $e_x$ is the function $e_x(x):=1$, $e_x(y):=0$ ($y\neq x$).
This is what makes this space important: For any Hilbert space $\calH$ with some orthonormal basis $X$, $\ell_2(X)$ is isomorphic to $\calH$. (And every Hilbert space has an orthonormal basis.\isalemmafn{orthonormal_basis_exists})
Since up to isomorphism, all Hilbert spaces are of the form $\ell_2(X)$, it often suffices to study the latter.
Due to this, we devote a separate theory \isathy{Complex_L2} to this Hilbert space.

\paragraph{Defining $\ell_2$.} To define $\ell_2(X)$ (which we will represent as a type \isatyp{'a ell2}) we first introduce a definition to abbreviate the condition $\sum_{x\in X}\abs{f(x)}^2<\infty$:
\begin{isabelletop}
definition ‹has_ell2_norm x ⟷ (λi. (x i)⇧2) abs_summable_on UNIV›[* for has_ell2_norm *]
\end{isabelletop}
\idxconst{has_ell2_norm}
Here \isaterm{f abs_summable_on UNIV} is defined in \isathy[HOL-Analysis]{Infinite_Sum} and means that the infinite sum $\sum_{x\in\isaterm{UNIV}}\norm{f(x)}$ exists\footnote{\label{footnote:infsum}%
  There are different definitions of infinite sums%
  \index{infinite sum}\index{sum!infinite}
  $\sum_{x\in X} f(x)$ for $f:A\to B$.
  \isathy[HOL-Analysis]{Infinite_Sum} (which is now included in the Isabelle distribution in \isasession{HOL-Analysis} but was originally developed as part of this formalization) uses a definition which is suitable when there is no canonical ordering on $X$.
  (As opposed to convergent series, where $X$ must be $\setN$ or something that can be identified with $\setN$, and where the order of summation matters.)
  If $B$ is a metric space with metric $d$,
  $\sum_{x\in X} f(x)$ is the unique value $s$ such that for all $\varepsilon>0$, there exists a finite set $F\subseteq X$ such that for all finite $G$ with $F\subseteq G\subseteq X$, we have $d\pb\paren{\sum_{x\in G} f(x), s}\leq\varepsilon$.
  The definition from \isathy[HOL-Analysis]{Infinite_Sum} is more general than this (the infinite sum is defined over any Hausdorff topological space) but we only need this special case in the present paper.} (recall that on complex numbers, norm and absolute value are the same).
And  similarly, for a square-summable function \isaterm{f} (i.e., satisfying \isaterm{has_ell2_norm f}), we can define its norm $\sqrt{\sum_{x\in X}\abs{f(x)}^2}$:
\begin{isabelletop}
definition ‹ell2_norm f = sqrt (∑⇩∞x. norm (f x)^2)›[* for ell2_norm *]
\end{isabelletop}
Here \isaterm{∑⇩∞[*x. x*]} refers to the infinite sum described in \autoref{footnote:infsum}.
\idxconst{ell2_norm}
Note that for \isaterm{f :: 'a ⇒ complex} where \isatyp{'a} is a finite type, the norm always exists, i.e., \isaterm{has_ell2_norm f} holds\isalemmafn{has_ell2_norm_finite} and is \isaterm{sqrt (∑x∈UNIV. (norm (f x))^2)}.\isalemmafn{ell2_norm_finite} (Note the different sum symbol: \isaterm{∑[*x∈UNIV. x*]} refers to Isabelle/HOL's easier-to-use finite sum \isaconstidx{Groups_Big.sum}, with no notion of convergence.)

Most importantly, we show that \isaterm{ell2_norm} is indeed a norm on the vectors satisfying \isaconst{has_ell2_norm}, namely it is non-negative,\isalemmafn{ell2_norm_geq0} non-degenerate,\isalemmafn{ell2_norm_0} commutes with scalar multiplication,\isalemmafn{ell2_norm_smult}, and satisfies the triangle inequality.\isalemmafn{ell2_norm_triangle}

\paragraph{The type \isatyp{ell2}.}
Since not all functions are square-summable, the type \isatyp{'a ⇒ complex} is not a normed vector space (with norm \isaconst{ell2_norm}).
Instead, we define the type \isatyp{'a ell2} that consists only of the square-summable functions \isatyp{'a ⇒ complex}:
\begin{isabelletop}
typedef 'a ell2 = ‹{f::'a⇒complex. has_ell2_norm f}›[* sorry *]
\end{isabelletop}
\idxtyp{ell2}%
We endow this type with pointwise addition and pointwise scalar multiplication, as well as \isaconst{has_ell2_norm} as the norm. Furthermore, we define the inner product of \isaterm{f}, \isaterm{g} as \isaterm{∑⇩∞x. cnj (f x) * g x} (where \isaconst{cnj} denotes complex conjugation). This sum always converges for square-summable functions.\isalemmafn{Infinite_Sum.abs_summable_product}
\begin{isabelletop}
lift_definition plus_ell2 :: ‹'a ell2 ⇒ 'a ell2 ⇒ 'a ell2›
  is ‹λf g x. f x + g x›[* sorry *]
lift_definition scaleC_ell2 :: ‹complex ⇒ 'a ell2 ⇒ 'a ell2›
  is ‹λc f x. c * f x›[* sorry *]
lift_definition norm_ell2 :: ‹'a ell2 ⇒ real› is ell2_norm[* sorry *]
lift_definition cinner_ell2 :: ‹'a ell2 ⇒ 'a ell2 ⇒ complex›
  is ‹λf g. ∑⇩∞x. cnj (f x) * g x›[* sorry *]
\end{isabelletop}
With these definitions (and the obvious ones for \isaconst{zero} etc.), we can show that \isatyp{ell2} is a Hilbert space, i.e., has type class \isaclass{chilbert_space}:
\begin{isabelletop}
instance ell2 :: (type) chilbert_space[* sorry *]
\end{isabelletop}
The Hilbert space \isatyp{'a ell2} will be the topic of the remainder of this section.

\paragraph{Kets (canonical basis).}
There is a canonical basis for the Hilbert space \isatyp{'a ell2} consisting of the functions $e_x$ with $e_x(x)=1$ and $e_x=0$ otherwise.
We define \isaterm{ket x} to be that function~$e_x$:
\begin{isabelletop}
lift_definition ket :: ‹'a ⇒ 'a ell2› is ‹λx y. if x=y then 1 else 0›[* sorry *]
\end{isabelletop}
\idxconst{ket}%
(The reason for the name ``ket''\index{ket} is that in quantum computing, we often write $\ket x$ for these vectors, pronounced ``ket x''.)
The kets form an orthonormal basis,\isalemmafn{is_onb_ket}
in particular two bounded operators on \isatyp{'a ell2} are equal when they coincide on all kets.\isalemmafn{equal_ket}
Also, \isaterm{ket x ∙⇩C ψ = Rep_ell2 ψ x}, where \isaterm{Rep_ell2 ψ x}\idxconst{Rep_ell2} refers to \isaterm{ψ(x)} when seeing \isaterm{ψ :: 'a ell2} as a function \isatyp{'a ⇒ complex}.\isalemmafn{cinner_ket_left}
Thus two $\elltwo$-vectors are equal if \isaterm{ket i ∙⇩C ψ = ket i ∙⇩C φ} for all \isaterm{i}.\isalemmafn{cinner_ket_eqI}

\paragraph{Truncated vectors.}%
\index{truncated vector (in $\elltwo$)}
It can be helpful to consider vectors \isatyp{'a ell2} that are non-zero only on a restricted (e.g., finite) set of locations.
If we think of square-summable \emph{sequences}, then for example $f_i(j):=f(j)$ for $j\leq i$ and $f_i(j):=0$ for $j>i$ gives us a finite ``truncated'' version $f_i$ of the sequence $f$.
The advantage of this is that $f_i\to f$ (with respect to the $\ell_2$-norm),\isalemmafn{trunc_ell2_lim_seq} so we can (for example) show facts about the simpler $f_i$ first, and then carry them over to $f$ via the limit $f_i\to f$.
However, in our theory we consider the more general case of square-summable functions, so this definition does not work.
(There is no general notion of a finite prefix.)
Instead, we define \isaterm{trunc_ell2 S ψ} to be \isaterm{ψ} but with all coefficients outside the set \isaterm{S} set to \isaterm{0}.
That is, \isaterm{Rep_ell2 (trunc_ell2 S ψ) x = (if x ∈ S then Rep_ell2 ψ x else 0)}.\isalemmafn{trunc_ell2.rep_eq}.
For finite \isaterm{S}, \isaterm{trunc_ell2 S ψ} then has only finitely many non-zero coefficients.
(In particular, it is in the (regular, i.e., not closed) span of the kets, \isaterm{cspan (range ket)}.\isalemmafn{trunc_ell2_ket_cspan})

We can then recover the limit property from the sequence example.
Namely, \isaterm{trunc_ell2 S ψ} (for finite sets \isaterm{S}) converges to \isaterm{ψ} when \isaterm{S} goes to \isaabbrev{UNIV} (the set containing the whole type). In Isabelle notation:
\begin{isabelletop}
lemma trunc_ell2_lim_at_UNIV:
  ‹((λS. trunc_ell2 S ψ) ⤏ ψ) (finite_subsets_at_top UNIV)›[* sorry *]
\end{isabelletop}
\idxlemma{trunc_ell2_lim_at_UNIV}%
This is interpreted as a limit of a \emph{net}, where the net is indexed by finite subsets of \isaabbrev{UNIV}, ordered by inclusion.
The Isabelle notation \isaterm{finite_subsets_at_top UNIV} indicates this.
(We have a more general version of this lemma, too, where \isaterm{trunc_ell2 M ψ} converges to \isaterm{trunc_ell2 N ψ} with respect to more general choices of the underlying directed set.\isalemmafn{trunc_ell2_lim_general})
The special case of sequences we mentioned above is an immediate consequence of this.\isalemmafn{trunc_ell2_lim_seq}

Besides this, we collect a number of facts about \isaconst{trunc_ell2},
e.g., that \isaconst{trunc_ell2} reduces the norm of \isaterm{ψ},\isalemmafn{trunc_ell2_reduces_norm}
that it is bounded linear (for fixed \isaterm{S}),\isalemmafn{bounded_clinear_trunc_ell2}
and various simple properties related to the interaction of \isaconst{trunc_ell2} with operations such as set-union, addition, etc.

\paragraph{Rank-1 operators (ctd.).}
\index{rank-1 operator!(over $\elltwo$)}%
\index{operator!rank-1 (over $\elltwo$)}
We introduced the concepts of rank-1 operators and butterflies on \autopageref{page:rank1}.
Specifically for butterflies of kets%
\index{butterfly!of kets}
(i.e., \isaterm{butterfly (ket x) (ket y)}, often written $\ket x\bra y$, hence the name butterfly),
we have the following facts: The set of all operators $\ket x\bra y$ is linearly independent.\isalemmafn{cindependent_butterfly_ket}
In the case of finite-dimensional spaces (i.e., over \isatyp{'a ell2} with \isatyp{'a} a finite type), they span the whole space.\isalemmafn{cspan_butterfly_ket}
(Note that this is not true in the infinite-dimensional case: The closed span of the butterflies are the compact operators, not the bounded operators.)
And also in the finite-dimensional case, the sum of all $\ket x\bra x$ (\isaterm{butterfly (ket x) (ket x)}) sum to the identity,\isalemmafn{sum_butterfly_ket}.
(The latter can be useful for introducing an identity in an expression, replacing it by this sum, and then using that terms involving butterflies can often be simplified a lot.)

\paragraph{Explicit operators.}%
\pagelabel{page:explicit_cblinfun}%
\index{explicit operators}%
\index{operator!explicit}
Bounded operators can be specified by describing them as a function, i.e., by telling which vector they map to which.
However, in the finite-dimensional cases, it is very common to describe a linear operator $a$ as a matrix%
\index{matrix!(explicit operator)}
$M$.
That is, $M_{ij}$ is the $i$-th coefficient of the image under $a$ of the $j$-th basis vector.
This idea generalizes to the infinite-dimensional case when describing operators over \isatyp{ell2} because \isatyp{ell2} comes with a canonical basis.
Specifically, to describe \isaterm{a :: 'a ell2 ⇒⇩C⇩L 'b ell2}, we give an infinite matrix \isaterm{M :: 'b ⇒ 'a ⇒ complex} so that \isaterm{M i j} is the \isaterm{i}-component of \isaterm{a *⇩V ket j}.
We introduce the operation \isaconstidx{explicit_cblinfun} that maps \isaterm{M} to \isaterm{a}.
That is, we have
\begin{isabelletop}
lemma Rep_ell2_explicit_cblinfun_ket[simp]:
  ‹Rep_ell2 (explicit_cblinfun M *⇩V ket a) b = M b a›
    if ‹explicit_cblinfun_exists M›[* sorry *]
\end{isabelletop}
\idxlemma{Rep_ell2_explicit_cblinfun_ket}%
Note that we have the condition \isaterm{explicit_cblinfun_exists M}\idxconst{explicit_cblinfun_exists} because not every matrix \isaterm{M} corresponds to a \emph{bounded} operator.
(Cf.~also \isaconst{cblinfun_extension_exists}, \autopageref{page:cblinfun_extension}.)
If \isatyp{'a},\isatyp{'b} are finite, then it exists,\isalemmafn{explicit_cblinfun_exists_finite_dim}.
Otherwise, we need to prove the existence, e.g., using this criterion:
\begin{isabelletop}
lemma explicit_cblinfun_exists_bounded:
  assumes ‹⋀S T ψ. finite S ⟹ finite T ⟹ (⋀a. a∉T ⟹ ψ a = 0) ⟹
            (∑b∈S. (cmod (∑a∈T. ψ a *⇩C M b a))⇧2)
                 ≤ B * (∑a∈T. (cmod (ψ a))⇧2)›
  shows ‹explicit_cblinfun_exists M›[* sorry *]
\end{isabelletop}
(\isaabbrevidx{cmod} is the absolute value of a complex number.)
In other words, we need to show that every finite submatrix is bounded (by the same bound \isaterm{B}).

Note that in the special case of 0/1-matrices, \isaconst{classical_operator} from the next section may be an easier to use alternative.

\paragraph{Classical operators.}%
\pagelabel{page:classical_operator}%
\index{classical operator}%
\index{operator!classical}
A common class of operators on \isatyp{'a ell2} are operators that map every \isaterm{ket x} to some \isaterm{ket (π x)} or to \isaterm{0}.
For example, in quantum computing, a classical function \isaterm{f} is often represented as a unitary that maps \isaterm{ket (x, y)} to \isaterm{ket (x, y + f x)}.
(This is why we call this a ``classical operator''; in quantum computation/information, it often corresponds to an operation that is in some way classical.)
Or the projector onto the closed span of all \isaterm{ket x} with \isaterm{x ∈ S} can equivalently be specified as the operator that maps \isaterm{ket x} to \isaterm{ket x} when \isaterm{x ∈ S} and to \isaterm{0} otherwise.
Such operators, can, of course, be defined using \isaconst{cblinfun_extension} (see \autopageref{page:cblinfun_extension}) or \isaconst{explicit_cblinfun} (see above).
But in that case, one needs to prove the existence of the operator (predicates \isaconst{cblinfun_extension_exists} or \isaconst{explicit_cblinfun_exists}) which is a tedious technical requirement.
Therefore we introduce a constant \isaconstidx{classical_operator} for the special case of classical operators.
The actual definition (in terms of \isaconst{cblinfun_extension}) is a bit technical, we omit it here an instead present its defining property:
\begin{isabelletop}
lemma classical_operator_ket:
  assumes ‹classical_operator_exists π›
  shows ‹classical_operator π *⇩V ket x
               = (case π x of Some i ⇒ ket i | None ⇒ 0)›[* sorry *]
\end{isabelletop}
That is, if \isaterm{π} is a partial function from \isatyp{'a} to \isatyp{'b} (in Isabelle: \isaterm{π :: 'a ⇒ 'b option}), then \isaterm{classical_operator π} is the operator that maps \isaterm{ket x} to \isaterm{ket i} when \isaterm{π x = Some i} (roughly speaking, it maps \isaterm{ket x} to \isaterm{ket (π x)} when \isaterm{π x} is defined), and maps \isaterm{ket x} to \isaterm{0} when \isaterm{π x = None} (when \isaterm{π x} is undefined).

This allows us, for example, to encode the two examples above as:
\begin{gather*}
\textnormal{\isaterm{classical_operator (λ(x,y). Some (x, y + f x))}}\\
\textnormal{\isaterm{classical_operator (λx. if x ∈ S then Some x else None)}}
\end{gather*}

Of course, we are still left with the requirement to prove that the classical operator exists.
For example, there exists no bounded operator that maps all \isaterm{ket x} to \isaterm{ket 0} (if \isaterm{x} ranges over an infinite type).\isalemmafn{bounded_extension_counterexample_2}
This is why \isalemma{classical_operator_ket} contains the premise \isaterm{classical_operator_exists π}.\idxconst{classical_operator_exists}
The big difference to \isaconst{cblinfun_extension_exists} or \isaconst{explicit_cblinfun_exists} is that \isaterm{classical_operator_exists π} is very easy to show in many cases.
Specifically, if \isaterm{π} is injective as a partial function (in Isabelle-notation: \isaterm{inj_map π}), then \isaterm{classical_operator_exists π},\isalemmafn{classical_operator_exists_inj} and \isaterm{classical_operator π} has norm \isaterm{[* x *]≤ 1}.\isalemmafn{classical_operator_norm_inj}

Moreover, if \isaterm{π} is total and injective or bijective, then \isaterm{classical_operator π} is an isometry\isalemmafn{isometry_classical_operator} or unitary,\isalemmafn{unitary_classical_operator} respectively.
Additionally, if a classical operator exists for both \isaterm{π} and \isaterm{ρ},
then it also exists for their composition \isaterm{π ∘⇩m ρ},\isalemmafn{classical_operator_exists_comp}
and the resulting classical operator is the composition \isaterm{classical_operator π o⇩C⇩L classical_operator ρ}.\isalemmafn{classical_operator_mult}.
Of course, the classical operator also always exists over \isatyp{'a ell2} for finite \isatyp{'a}.\isalemmafn{classical_operator_exists_finite}
And lastly, it is worth mentioning that the adjoint of \isaterm{classical_operator π} is \isaterm{classical_operator (inv_map π)} for injective (not necessarily total)~\isaterm{π}.\isalemmafn{classical_operator_adjoint}
(Here \isaterm{inv_map π} is the inverse of the partial function~\isaterm{π}.)


\section{Matrices and bounded operators}
\label{sec:matrix}

Our development targets bounded linear operators over finite- or infinite-dimensional complex vector spaces.
In the finite dimensional case, linear operators are essentially matrices.
One extensive Isabelle/HOL library dealing with matrices is the \isasessionidx{Jordan_Normal_Form} library~\cite{Jordan_Normal_Form-AFP}, short JNF.
It provides a modeling of matrices over arbitrary fields (including complex numbers).
One limitation of the JNF library compared to our framework is its lack of type safety (one cannot, e.g., define a type of $2\times2$ matrices, matrices of all sizes inhabit the same type).
But the strength of the JNF library lies in its extensive support for numeric algorithms (such as the computation of the eponymous Jordan normal form), and its support for code generation:
Given a description of a matrix in Isabelle, the JNF library can compute the results of the various algorithms directly using Isabelle's code generation mechanism \cite{codegen}, starting from simple operations such as multiplication of matrices to complex ones such as the computation of eigenvectors, Jordan normal forms, etc.
In principle, those algorithms could be implemented in our library as well for finite-dimensional operators.
However, to reduce duplication and make use of the rich existing development, we chose another way:

In the theory \isathy{Cblinfun_Matrix}, we formalize the correspondence between bounded operators on finite-dimensional spaces (type \isatyp{cblinfun}) and JNF's \emph{complex} matrices (type \isatyp{complex mat}).
We then connect most of the operations on vectors, matrices, and subspaces with corresponding concepts in JNF;
this allows then to translate computable problems to JNF and solve them there using the generated code.
(And in \autoref{sec:code} below, we use this to directly support code generation on finite-dimensional bounded operators.)

\paragraph{Correspondence between JNF-vectors and our vectors.}
In JNF, a vector\index{vector!(in \isasession{Jordan_Normal_Form})} is an element of a type \isatyp{complex vec}\idxtyp{vec}\idxtyp{complex vec};
modeling an $n$-tuple of complex numbers.
Since $n$-vectors for all finite~$n$ are in the same type \isatyp{complex vec}, we get the length of the vector \isaterm{n} as \isaterm{dim_vec v}\idxconst{dim_vec}.
And the $i$-th component is \isaterm{vec_index v i}\idxconst{vec_index} or \isatermpre{unbundle jnf_syntax}{v $ i}.
Now, as we see, the addressing of the components of the vector is via natural numbers.
In contrast, in our development, a vector space does of course have a basis, but there is no canonical identification between the basis vectors and the naturals $1,\dots,n$ in general.
For this purpose, we introduced the type class \isaclass{basis_enum} (see \autopageref{page:basis_enum}).
It provides the constant \isaconst{canonical_basis} which enumerates a finite basis $[b_1,\dots,b_n]$.
For a JNF vector \isaterm{v}, we define \isaterm{basis_enum_of_vec v :: 'a[*::basis_enum*]}\idxconst{basis_enum_of_vec} as the vector
$\sum_{i=1}^n (\isatermpre{unbundle jnf_syntax}{v $ i})\cdot  b_i$. 
E.g., \isaterm{basis_enum_of_vec v :: bool ell2} will expect \isaterm{v} to have dimension \isaterm{dim_vec v = 2}
and return \isatermpre{unbundle jnf_syntax}{(v $ 0) *⇩C ket False + (v $ 1) *⇩C ket True}.
We also define the inverse \isaterm{vec_of_basis_enum :: 'a[*::basis_enum*] ⇒ complex vec}\idxconst{vec_of_basis_enum}.
We have \isaterm{basis_enum_of_vec (vec_of_basis_enum ψ) = ψ},\isalemmafn{vec_of_basis_enum_inverse}
and \isaterm{vec_of_basis_enum (basis_enum_of_vec v) = v} assuming \isaterm{dim_vec v} is the dimension of the space containing \isaterm{v}.\isalemmafn{basis_enum_of_vec_inverse}

These two functions allow us to formulate the correspondence between operations on vectors in our formalization (if they live in a type that has been instantiated with class \isaclass{basis_enum}) and JNF-vectors.

\paragraph{Operations on vectors.}
For many operations on vectors in our library (i.e., on types \isatyp{'a :: complex_vector}, there is a corresponding operation on JNF vectors (type \isatyp{complex vec}).
For example, we show:
\begin{isabelletop}
lemma vec_of_basis_enum_add:
  ‹vec_of_basis_enum (a + b) = vec_of_basis_enum a + vec_of_basis_enum b›[*sorry*]
\end{isabelletop}
\idxlemma{vec_of_basis_enum_add}%
and in the opposite direction
\begin{isabelletop}
lemma basis_enum_of_vec_add:
  assumes ‹dim_vec v1 = length (canonical_basis :: 'a[*::basis_enum*] list)›
          ‹dim_vec v2 = length (canonical_basis :: 'a list)›
  shows ‹(basis_enum_of_vec (v1 + v2) :: 'a)
                    = basis_enum_of_vec v1 + basis_enum_of_vec v2›[* sorry *]
\end{isabelletop}
\idxlemma{basis_enum_of_vec_add}%
The premise is necessary because \isaterm{v1}, \isaterm{v2} must be vectors of the correct length for \isaconst{basis_enum_of_vec} to be meaningful.
We show analogous correspondence lemmas for \isaconst{scaleC},\isalemmafn{vec_of_basis_enum_scaleC}\fnsep\isalemmafn{basis_enum_of_vec_mult} \isaconst{minus},\isalemmafn{vec_of_basis_enum_minus} \isaconst{cinner},\isalemmafn{cscalar_prod_vec_of_basis_enum}\fnsep\isalemmafn{cinner_basis_enum_of_vec} \isaconst{norm}\isalemmafn{norm_vec_of_basis_enum} (for the latter, we first define the norm \isaterm{norm_vec} on JNF vectors), elements of the \isaconst{canonical_basis},\isalemmafn{vec_of_basis_enum_canonical_basis}\fnsep\isalemmafn{basis_enum_of_vec_unit_vec} \isaterm{0},\isalemmafn{vec_of_basis_enum_zero} \isaconst{cspan},\isalemmafn{complex_vec_space.vec_of_basis_enum_cspan} \isaconst{is_ortho_set}.\isalemmafn{corthogonal_vec_of_basis_enum}
For \isatyp{'a :: one_dim} additionally: \isaterm{times},\isalemmafn{vec_of_basis_enum_times} \isaconst{inverse},\isalemmafn{vec_of_basis_enum_to_inverse} \isaconst{divide},\isalemmafn{vec_of_basis_enum_divide} \isaterm{1}.\isalemmafn{vec_of_basis_enum_1}

All operations on the JNF-side can be explicitly computed using Isabelle code generation.
(This also holds for the operations described in the remainder of this section.)

\paragraph{Correspondence between JNF-matrices and our bounded operators.}
In JNF, a matrix%
\index{matrix!(in \isasession{Jordan_Normal_Form})}
is an element of a type \isatyp{complex mat}\idxtyp{mat}; modeling an $n\times m$ complex matrix.
Since the $n\times m$ matrices for all finite $n,m$ are in the same type \isatyp{complex vec}, we get the number of rows/columns $n,m$ of a matrix \isaterm{M} as \isaterm{dim_row M}\idxconst{dim_row} and \isaterm{dim_col M}\idxconst{dim_col}.
And the $(i,j)$-component is \isaterm{index_mat M (i,j)} or \isatermpre{unbundle jnf_syntax}{M $$ (i,j)}.
A matrix \isaterm{M :: complex mat} is applied to a vector \isaterm{v :: complex vec} by writing \isaterm{mult_mat_vec M v} or \isatermpre{unbundle jnf_syntax}{M *⇩v v}.
(Note the small
\isatermpre{consts dummy :: nat ("⇩v")}{⇩v}, compared with our operation \isaterm{[*x*]*⇩V[*y*]}.)

Analogously to the case of vectors, we can also identify matrices with bounded operators \isatyp{'a ⇒⇩C⇩L 'b}, as long as \isatyp{'a}, \isatyp{'b} are of class \isaclass{basis_enum}.
Namely, \isaterm{cblinfun_of_mat M}\idxconst{cblinfun_of_mat} is the bounded operator mapping \isaterm{ψ} to \isatermpre{unbundle jnf_syntax}{basis_enum_of_vec (M *⇩v vec_of_basis_enum ψ)}, assuming \isaterm{M} is a matrix of the right dimension.
And \isaterm{mat_of_cblinfun} is the inverse operation.

We have \isaterm{cblinfun_of_mat (mat_of_cblinfun a) = a},\isalemmafn{mat_of_cblinfun_inverse}
and \isaterm{mat_of_cblinfun (cblinfun_of_mat M) = M} assuming \isaterm{M} has the right size.\isalemmafn{cblinfun_of_mat_inverse}

\paragraph{Operations on matrices.}
Analogously to the vectors, we have lemmas relating operations on bounded operators to operations on JNF-matrices.
For example:
\begin{isabelletop}
[* unbundle jnf_syntax *]lemma mat_of_cblinfun_cblinfun_apply:
  ‹vec_of_basis_enum (F *⇩V u) = mat_of_cblinfun F *⇩v vec_of_basis_enum u›[* sorry *]
\end{isabelletop}
\idxlemma{mat_of_cblinfun_cblinfun_apply}%
relating application of bounded operators to vectors (\isaconst{cblinfun_apply}, \isaterm{[*x*]*⇩V[*y*]}) with matrix-vector multiplication in JNF (\isaconst{mult_mat_vec}, \isatermpre{unbundle jnf_syntax}{[*x*]*⇩v[*y*]}).
And of course the opposite direction (\isatermpre{unbundle jnf_syntax}{basis_enum_of_vec (M *⇩v x) = [*tmp*][*!\dots*]}).\isalemmafn{basis_enum_of_vec_cblinfun_apply}
We show analogous correspondence lemmas for
\isaconst{cblinfun_compose},\isalemmafn{mat_of_cblinfun_compose}\fnsep\isalemmafn{cblinfun_of_mat_times} \isaconst{scaleC},\isalemmafn{mat_of_cblinfun_scaleC} \isaconst{plus},\isalemmafn{mat_of_cblinfun_plus}\fnsep\isalemmafn{cblinfun_of_mat_plus}~\isaterm{0},\isalemmafn{mat_of_cblinfun_zero} \isaconst{id_cblinfun},\isalemmafn{mat_of_cblinfun_id} \isaconst{uminus},\isalemmafn{mat_of_cblinfun_uminus}\fnsep\isalemmafn{cblinfun_of_mat_uminus} \isaconst{minus},\isalemmafn{mat_of_cblinfun_minus}\fnsep\isalemmafn{cblinfun_of_mat_minus} \isaconst{adj},\isalemmafn{mat_of_cblinfun_adj}\fnsep\isalemmafn{cblinfun_of_mat_adjoint} \isaconst{explicit_cblinfun},\isalemmafn{mat_of_cblinfun_explicit_cblinfun} \isaconst{classical_operator},\isalemmafn{mat_of_cblinfun_classical_operator} \isaconst{vector_to_cblinfun},\isalemmafn{mat_of_cblinfun_vector_to_cblinfun} \isaabbrev{proj}.\isalemmafn{mat_of_cblinfun_proj}
And for \isatyp{'a :: one_dim}, \isatyp{'b :: one_dim}:
\isaconst{times}\isalemmafn{mat_of_cblinfun_times},
\isaconst{one_dim_iso}\isalemmafn{mat_of_cblinfun_one_dim_iso},
\isaterm{1}\isalemmafn{mat_of_cblinfun_1} ($1\times 1$ identity matrix).

\paragraph{Operations on subspaces.}%
\index{subspace!(representation in JNF)}
In our library, we also have the type of (closed) subspaces (type \isatyp{ccsubspace}).
There is no comparable type in JNF, but we can describe a subspace by giving generating set of vectors.
That is, in JNF we can represent a subspace as type \isatyp{complex vec list}; we will call these JNF-subspaces.
Then given a JNF-subspace \isaterm{S :: complex vec list},
the corresponding subspace in our library is then \isaterm{ccspan (set (map basis_enum_of_vec S))}.
(We do not provide a map in the opposite direction, i.e., from \isatyp{'a ccsubspace} to \isatyp{complex vec list}, since the relationship is not one-to-one.)

Then we can relate operations on subspaces to corresponding operations on lists of JNF-vectors. For example:
\begin{isabelletop}
lemma ccspan_leq_using_vec: ‹(ccspan (set A) ≤ ccspan (set B)) ⟷
    is_subspace_of_vec_list (length (canonical_basis))
              (map vec_of_basis_enum A) (map vec_of_basis_enum B)›[* sorry *]
\end{isabelletop}
\idxlemma{ccspan_leq_using_vec}%
Here \isaconstidx{is_subspace_of_vec_list} checks whether its second argument is contained in the span of its first.
This constant is provided by our library; no such operation is provided in JNF.
In a nutshell, it works by doing a Gram--Schmidt orthogonalization%
\index{Gram--Schmidt orthogonalization}
of the second argument and then using that to check if all vectors in the first argument are in the span.
JNF provides a Gram--Schmidt orthogonalization procedure \isaconstidx{gram_schmidt} but it is not suitable for our purposes because it requires its argument to consist of \emph{linearly independent} vectors.
Relying heavily on existing code in JNF, we provide a tweaked variant of Gram--Schmidt (\isaconstidx{gram_schmidt0}) that does not have this restriction.

We use the same Gram--Schmidt procedure to relate the operator \isaterm{Proj S} to a corresponding matrix in JNF.\isalemmafn{mat_of_cblinfun_Proj_ccspan}
(Using that projector onto \isaterm{S} can be written as $\sum_i \psi_i\adj{\psi_i}$ if the $\psi_i$ form an orthonormal basis of \isaterm{S}.)
Finally, we also provide JNF-code for \isaconst{cblinfun_image}.\isalemmafn{cblinfun_image_ccspan_using_vec}

\section{Code generation}
\label{sec:code}
\index{code generation}

Code generation \cite{codegen} is a mechanism in Isabelle/HOL to transform terms (e.g., of type \isatyp{bool} or \isatyp{nat}) into an executable program (in various languages such as Standard ML, Ocaml, Haskell, and Scala).
The resulting program can then be used as verified code outside of Isabelle, or it can be used directly by Isabelle to prove a subgoal by explicit computation.
For example:
\begin{isabelletop}
lemma ‹(1234::nat) * 4321 = 5332114›
  by eval
\end{isabelletop}
Here the \isatop{[*lemma True by *]eval} method internally produces code that computes and compares the lhs and rhs, runs it, and discharges the current goal if the computation returns \isaconst{True}.
Note that proving this theorem manually (by using the various elementary lemmas about multiplication, addition, and binary representations of integers) would be very cumbersome.

In order to generate the code, Isabelle needs to know how to generate code for every operation occurring in the term to be computed.
(Here, code for \isaterm{[*1*]*[*2*]}, but also less obviously code for ``computing'' the integer literals \isaterm{1234}, \isaterm{4321}, and \isaterm{5332114} for their internal representation.)
This is configured by registering lemmas of a specific form with the code generation subsystem.
For example, recursive code for computing the gcd on integers is automatically configured by providing the lemma \isaprop{gcd k l = ¦if l = 0 then k else gcd l (¦k¦ mod ¦l¦)¦},\isalemmafn{GCD.gcd_code_int}.
Since the lemmas passed to the code generation have to be proven first, this ensures that we do not add any incorrect rules that make proving via evaluation unsound.%
\footnote{However, this mechanism is not perfect.
  First, using code generation uses a much larger trusted core than the normal reasoning in Isabelle.
  Second, it is still possible to misconfigure the code generation and produce incorrect code by using commands such as \isatop{code_printing[* constant plus => (SML) infix 1 "+"*]}.}

For a deeper understanding of the code generation mechanism, we refer to \cite{codegen}.

In the theory \isathyidx{Cblinfun_Code}, we set up code generation for computing with \emph{finite-dimensional} vectors and operators.
This then allows us to automatically evaluate formulas and prove facts involving explicitly given bounded operators and vectors.
Since finite-dimensional operators and vectors stand in one-to-one correspondence with JNF-matrices and -vectors (see \autoref{sec:matrix}), and since JNF already provides extensive support for code generation, we do not reinvent the wheel but instead create shallow wrappers around the JNF-code generation.
From the mathematical point of view, most of the work was already done in the theory \isathy{Cblinfun_Matrix};
here we only need to ``explain'' to Isabelle's code generation mechanism how to use those results.

\paragraph{Bounded operators.}%
\index{bounded operator!(code generation)}
To represent operators as matrices in generated code, we register the type \isatyp{cblinfun}%
\idxtyp[(code generation)]{cblinfun}
as an ``abstract type''%
\index{abstract type!(code generation)}\index{type!abstract (code generation)}
based on JNF-matrices.
To do so, we need to provide an embedding from bounded operators into JNF-matrices (\isaconst{mat_of_cblinfun} from \isathy{Cblinfun_Matrix}), and its inverse \isaconst{cblinfun_of_mat}.
We tell the code generation that they are inverses and that we want to use them to define an abstract type by:
\begin{isabelletop}
declare mat_of_cblinfun_inverse [code abstype]
\end{isabelletop}
(Recall that \isalemma{mat_of_cblinfun_inverse} states \isaterm{cblinfun_of_mat (mat_of_cblinfun B) = B} for finite-dimensional%
\footnote{Strictly speaking, we require a bounded operator of type \isatyp{'a ⇒⇩C⇩L 'b} where \isatyp{'a}, \isatyp{'b} have type class \isaclass{basis_enum}.}
\isaterm{B}.)
Isabelle automatically extracts all relevant information from the statement of that theorem.
Note that this does not require that every JNF-matrix corresponds to a valid bounded operator.
This is important because the type \isatyp{complex vec} of JNF-matrices contains matrices of all sizes in a single type.

Registering such an abstract type tells the code generation to internally represent all bounded operators as \isaterm{cblinfun_of_mat M} for some JNF-matrix \isaterm{M}.
More precisely, in the generated code (e.g., OCaml code), there will be type with constructor \textbf{Cblinfun\_of\_mat}, and \textbf{Cblinfun\_of\_mat} \isaterm{M} will be interpreted as meaning \isaterm{cblinfun_of_mat M}.
(And \isaterm{M} in turn will be represented with whatever OCaml representation was configured by JNF.)
This way, all finite-dimensional operators can be represented as matrices in code. (Even though we have originally defined operators as functions between vector spaces.)

Next, we need to tell Isabelle for each operation on bounded operators how to compute it. For this purpose, we have to specify lemmas of the form \isaterm{mat_of_cblinfun X = Y} where \isaterm{X} is the kind of expression we wish to compute, and \isaterm{Y} the recipe how to compute it.
For example, \isalemma{mat_of_cblinfun_plus} from \isathy{Cblinfun_Matrix} says \isaterm{mat_of_cblinfun (F + G) = mat_of_cblinfun F + mat_of_cblinfun G}, and we declare it as a code equation using
\begin{isabelletop}
declare mat_of_cblinfun_plus [code]
\end{isabelletop}
Then the code generation knows that adding \textbf{Cblinfun\_of\_mat} \isaterm{M} and
 \textbf{Cblinfun\_of\_mat} \isaterm{N} can be done my first computing \isaterm{M+N} using JNF's code and then returning \textbf{Cblinfun\_of\_mat} \isaterm{(M+N)}.)

Similarly, we declare code generation rules for the operations and constants \isaconst{minus} (\isaterm{[*0*]-[*0*]}), \isaconst{id_cblinfun}, \isaterm{0}, \isaconst{uminus} (unary \isaterm{[*0*]-[*0*]}), \isaconst{classical_operator}, \isaconst{cblinfun_compose} (\isaterm{[*F*]o⇩C⇩L[*G*]}), \isaconst{scaleC} (\isaterm{[*x*]*⇩C[*y*]}), \isaconst{scaleR} (\isaterm{[*x*]*⇩R[*y*]}), \isaconst{adj}, \isaterm{1} (on one-dimensional spaces, see \autoref{sec:one-dim}), \isaconst{explicit_cblinfun}, \isaconst{equal} (equality tests).

\paragraph{Vectors.}%
\index{vector!(code generation)}
For operators, we declared an ``abstract type'' to support code generation.
Ideally, we would like to do the same, i.e., declare an abstract type by providing the fact \isaterm{basis_enum_of_vec (vec_of_basis_enum ψ) = ψ} where \isaterm{ψ} has type \isatyp{'a :: basis_enum} (essentially arbitrary finite-dimensional complex vector spaces).

Unfortunately, this does not work; the type \isatyp{'a :: basis_enum}\idxclass[(code generation)]{basis_enum} is too general.
The code generation mechanism only allows us to set up abstract types that involve a type constructor.
(E.g., \isatyp{'a list}, \isatyp{'a ell2}, or similar.)
Our solution is to define an abstract type only for the specific type \isatyp{'a ell2}%
\idxtyp[(code generation)]{ell2}
(see \autoref{sec:ell2}), not for all complex vector spaces.
We do this by defining constants \isaterm{ell2_of_vec :: complex vec ⇒ ('a::enum) ell2} and \isaterm{vec_of_ell2 :: ('a::enum) ell2 ⇒ complex vec} that map between \isatyp{ell2}-vectors and JNF-vectors.
This is quite trivial because we can simply define them to be identical to the functions \isaconst{basis_enum_of_vec} and \isaconst{vec_of_basis_enum} from \isathy{Cblinfun_Matrix}, only with a narrower type.
We can then declare the abstract type of \isaterm{ell2}-vector via:
\begin{isabelletop}
lemma vec_of_ell2_inverse [code abstype]: ‹ell2_of_vec (vec_of_ell2 B) = B›[* sorry *]
\end{isabelletop}
\idxlemma{vec_of_ell2_inverse}

And then, as for operators, we can declare the rules for the different operations on vectors, e.g.,
\begin{isabelletop}
lemma ell2_of_vec_minus[code]:
  ‹vec_of_ell2 (x - y) = (vec_of_ell2 x) - (vec_of_ell2 y)›[* sorry *]
\end{isabelletop}
This follows trivially from \isalemma{vec_of_basis_enum_minus} from \isathy{Cblinfun_Matrix}.

And analogously for \isaconst{ket},
\isaconst{scaleC} (\isaterm{[*x*]*⇩C[*y*]}), \isaconst{scaleR} (\isaterm{[*x*]*⇩R[*y*]}),
\isaconst{plus} (\isaterm{[*0*]+[*0*]}), \isaterm{0}, \isaconst{uminus}~(unary \isaterm{[*0*]-[*0*]}), \isaconst{cinner}~(\isaterm{[*x*]∙⇩C[*y*]}),
\isaconst{norm}, and -- for one-dimensional spaces --
\isaconst{times} (\isaterm{[*x*]*[*y*]}), \isaconst{divide} (\isaterm{[*x*]/[*y*]}), \isaconst{inverse}, and~\isaterm{1}.

Note that while this approach gives us code generation only for the type \isatyp{'a ell2}, not the more general \isatyp{'a :: basis_enum}, the approach makes it very easy to add support for other types of vectors.
Say, we want to support some finite-dimensional type \isatyppre{typedecl 'a myvector}{'a myvector} of vectors.
Then we simply need to define \isatermpre{typedecl 'a myvector consts myvector_of_vec :: 'a}{myvector_of_vec :: complex vec ⇒ 'a myvector} and \isatermpre{typedecl 'a myvector consts vec_of_myvector :: 'a}{vec_of_myvector :: 'a myvector ⇒ complex vec}, declare an abstract type (with a lemma analogous to \isalemma{vec_of_ell2_inverse}), and declare all code generation rules.
This basically boils down to copying the corresponding lemmas for \isatyp{ell2}, and replacing all occurrences of \texttt{ell2} by \texttt{myvector}.
The proofs will still be trivial because they reduce directly to the corresponding theorems in \isathy{Cblinfun_Matrix}.

\paragraph{Combined matrix/vector operations.}
What is missing from the supported operations so far is \isaconst{cblinfun_apply}\pagelabel{page:cblinfun_apply-code} (\isaterm{[*x*]*⇩V[*y*]}), that applies an operator to a vector.
Here we face another problem:
We would like to add the rule
\isaterm{vec_of_ell2 (M *⇩V x) = mult_mat_vec (mat_of_cblinfun M) (vec_of_ell2 x)} to support \isaconst{cblinfun_apply} to the code generation setup.
However, this is rejected by Isabelle, presumably because the type of the constant \isaconst{cblinfun_apply} (\isaterm{[*x*]*⇩V[*y*]}) is wider than its use in that rule (where it is restricted to \isatyp{ell2}).\footnote{%
  When adding attribute \isaattrib{[code]} to this rule, Isabelle gives a warning saying ``\texttt{Projection as head in equation}'' and silently ignores the rule.}
The solution is to define a new constant \isaconst{cblinfun_apply_ell2}, equal to \isaconst{cblinfun_apply} but of a more restricted type \isatyp{'a ell2 ⇒⇩C⇩L 'b ell2 ⇒ 'a ell2 ⇒ 'b ell2}.
Code generation for this constant is then added by the rule \isaterm{vec_of_ell2 (cblinfun_apply_ell2 M x) = mult_mat_vec (mat_of_cblinfun M) (vec_of_ell2 x)}.\isalemmafn{cblinfun_apply_ell2}
And we can instruct the code generation to replace occurrences of \isaconst{cblinfun_apply} by \isaconst{cblinfun_apply_ell2} whenever possible in a preprocessing step by adding the attribute \isaattrib{[code_abbrev]} to the definition of \isaconst{cblinfun_apply_ell2}.

Another constant involving both vectors and operators is \isaconst{vector_to_cblinfun}.
We use the same trick to support this one.

\paragraph{Subspaces.}%
\index{subspace!(code generation)}
Mathematically, a subspace (type \isatyp{'a ccsubspace})%
\idxtyp[(code generation)]{ccsubspace}
is simply a set of vectors closed under linear combinations.%
\footnote{The type \isatyp{'a ccsubspace} actually refers to \emph{closed} subspaces.
  However, in the context of code generation, the vector spaces are finite dimensional, hence all subspaces are closed.
  So this distinction does not matter in the present context.}
Since these sets are infinite, we cannot directly represent them in code.
There are many possible representations of subspaces, e.g., images of operators, kernels of operators, via a basis, via a generating set, etc.
We chose the representation via a generating set since it allows a sparse representation (i.e., low-dimensional subspaces can be represented by few vectors),
and unlike a basis, it is not necessary to throw out linearly-dependent vectors after each computation step (though we may do so for efficiency).

More precisely, we represent a subspace as a \isatyp{complex vec list}, i.e., a list \isaterm{S} of JNF-vectors.
And the corresponding \isatyp{ccsubspace} is given by \isaterm{SPAN S}, where \isaconst{SPAN} is defined roughly as follows:%
\footnote{The actual definition is a little more complicated because it first removes all JNF-vectors from \isaterm{S} that do not have the correct dimension.
  This is because we want \isaconst{SPAN} to be well-defined even when \isaterm{S} contains some ``garbage'' vectors.
  Since such garbage vectors are never produced by any of the operations we define, this is merely a technicality.}
\begin{isabelletop}
definition ‹SPAN S = ccspan (basis_enum_of_vec ` set S)›[* for SPAN *]
\end{isabelletop}
(\isaterm{f ` A} is Isabelle notation for $f(A)$ and \isaconstidx{set} converts a list to a set.)
That is, \isaterm{SPAN S} is the span (\isaconst{ccspan}) of the vectors in \isaterm{S} (seen as a set), after translating them from JNF-vectors to our vectors using \isaconst{basis_enum_of_vec}.

Then we need to tell the code generation about this correspondence.
For operators and vectors, we did so by defining an ``abstract type'' in the parlance of the code generation mechanism.
However, the abstract type mechanism would require that each subspace has a unique representation;
this is not the case here because there are many possible ways to describe the same subspace by a generating set.
Instead, we can declare it as a ``code datatype''%
\index{code datatype}\index{datatype!code} using the command \isatop{code_datatype SPAN}.
Similar to the ``abstract type'', this relates the domain of \isaconst{SPAN} as the representation in the generated code with the codomain as the type used in the mathematical expressions we want to translate to code.
The code datatype does not require the representation to be unique.

With that setup, we can, for example, declare the code generation for the sum (least upper bound) of two subspaces.
The generated code simply concatenates the generating sets:
\begin{isabelletop}
lemma sup_spans[code]: ‹SPAN A ⊔ SPAN B = SPAN (A @ B)›[*sorry*]
\end{isabelletop}
(Recall that \isaterm{[*x*]⊔[*x*]} is Isabelle syntax for least upper bounds, and \isaterm{[*x*]@[*x*]} for list concatenation.)

We similarly provide code equations for subpaces for:
\isaconst{top}\isalemmafn{top_ccsubspace_code} (\isaterm{⊤}, the subspace containing everything), \isaconst{bot}\isalemmafn{bot_as_span} (\isaterm{⊥}, \isaterm{0}, the subspace containing only $0$), \isaconst{ccspan},\isalemmafn{span_Set_Monad}
\isaconst{equal}\isalemmafn{equal_ccsubspace_code} (equality test), \isaconst{less_eq}\isalemmafn{leq_ccsubspace_code} (\isaterm{[*x*]≤[*x*]}), \isaconst{cblinfun_image}\isalemmafn{cblinfun_image_code}\fnsep\isalemmafn{range_cblinfun_code} (\isaterm{[*a*]*⇩S[*s*]}),
\isaconst{uminus}\isalemmafn{uminus_Span_code} (unary \isaterm{-[*x*]}, orthogonal complement), \isaconst{kernel},\isalemmafn{kernel_code}
\isaconst{inf}\isalemmafn{inf_ccsubspace_code} (\isaterm{[*x*]⊓[*x*]}, intersection of subspaces),
\isaconst{Sup}\isalemmafn{Inf_ccsubspace_code} (sum of a family of subspaces),
\isaconst{Inf}\isalemmafn{Inf_ccsubspace_code} (intersection of a family of subspaces).

Many of these are implemented using Gram--Schmidt orthogonalization (see \autoref{sec:matrix}).
Also, some technical challenges similar to the one encountered with \isaconst{cblinfun_apply} (\autopageref{page:cblinfun_apply-code}) occur and are solved with similar ideas.
We omit the details here and refer the interested reader to the comments in the theory files.

\paragraph{Examples.}
After all this set-up, we can finally do things such as computing the norm of the result of projecting \isaterm{ket 0} onto the space spanned by \isaterm{ket 0 + ket 1}:\footnote{The examples assumes that the type \isatyp{bit} is suitably set up.
  For this, import the theories \isathylong{HOL-Library}{Z2}, \isathylong{HOL-Library}{Code_Cardinality} and
  include the code
\isatop{%
instantiation bit :: ‹{enum,card_UNIV}› begin
definition "enum_bit = [0::bit,1]"
definition "enum_all_bit P ⟷ P (0::bit) ∧ P 1"
definition "enum_ex_bit P ⟷ P (0::bit) ∨ P 1"
definition "finite_UNIV = Phantom(bit) True"
definition "card_UNIV = Phantom(bit) 2"
instance sorry (* Use actual proof in production. *)
end}}
The command
\begin{isabelletoppre}{instantiation bit :: ‹{enum,card_UNIV}› begin
definition "enum_bit = [0::bit,1]"
definition "enum_all_bit P ⟷ P (0::bit) ∧ P 1"
definition "enum_ex_bit P ⟷ P (0::bit) ∨ P 1"
definition "finite_UNIV = Phantom(bit) True"
definition "card_UNIV = Phantom(bit) 2"
instance sorry
end}
value ‹norm (Proj (ccspan {ket (0::bit) + ket 1}) *⇩V ket 0)›
\end{isabelletoppre}
prints
\isaterm{1 / 2 * sqrt 2}.

And
\begin{isabelletoppre}{instantiation bit :: ‹{enum,card_UNIV}› begin
definition "enum_bit = [0::bit,1]"
definition "enum_all_bit P ⟷ P (0::bit) ∧ P 1"
definition "enum_ex_bit P ⟷ P (0::bit) ∨ P 1"
definition "finite_UNIV = Phantom(bit) True"
definition "card_UNIV = Phantom(bit) 2"
instance sorry
end}
lemma ‹ccspan {ket 0} ≤ ccspan {ket 0 + ket 1, ket (1::bit)}›
  by eval
\end{isabelletoppre}
automatically shows the inclusion of one subspace automatically by computation (using Gram--Schmidt under the hood).


\paragraph{Acknowledgments.}
The research was supported by the ERC consolidator grant CerQuS (819317), the PRG team grant “Secure Quantum Technology” (PRG946) from the Estonian Research Council, and the Estonian Centre of Excellence in IT (EXCITE) funded by ERDF, and the Estonian Centre of Excellence, and the Estonian Centre of Excellence ``Foundations of the Universe'' (TK202) funded by the Estonian Ministry of Education and Research. We also thank the anonymous reviewers from ITP 2026 for valuable feedback.

\makeatletter
\def\blx@bibheading#1#2{
  \section*{References}
  \addcontentsline{toc}{section}{References}
}

\printbibliography

\renewcommand\symbolindexentry[4]{
  \noindent\hbox{\hbox to 2in{$#2$\hfill}\parbox[t]{3.5in}{#3}\hbox to 1cm{\hfill #4}}\\[2pt]}

\renewenvironment{thesymbolindex}{%
  \section*{Symbol index}%
  \addcontentsline{toc}{section}{Symbol index}%
  \begin{center}}{\end{center}}

\printsymbolindex

\renewenvironment{theindex}{
  \section*{Keyword index}\addcontentsline{toc}{section}{Keyword index}%
  \setlength{\multicolsep}{0pt}%
  \begin{multicols}{2}
    \raggedright
    \setlength{\parindent}{0pt}%
    \let\item\@idxitem
  }{
  \end{multicols}
}

\printindex

\end{document}